\documentclass[aps,prd,preprintnumbers,nofootinbib,preprint]{revtex4}

\usepackage{graphicx}
\usepackage{amsmath}
\usepackage{amssymb}
\usepackage{color}
\usepackage{soul}

\def\be{\begin{equation}}
\def\ee{\end{equation}}
\def\bea{\begin{eqnarray}}
\def\eea{\end{eqnarray}}

\def\afbt{A_{FB}^t}
\def\afbl{A_{FB}^\ell}
\def\coth{{\text{coth}}}
\def\csch{{\text{csch}}}

\begin{document}

\preprint{ANL-HEP-PR-12-70, IIT-CAPP-12-09}

\title{Interpretations and Implications of  the Top Quark\\ Rapidity Asymmetries $A_{FB}^t$ 
and $A_{FB}^{\ell}$}

\author{Edmond L. Berger$^{\dagger}$, Qing-Hong Cao$^{*}$, Chuan-Ren Chen$^{\dagger,\S}$,
Hao Zhang$^{\dagger,\$}$}
\affiliation{
\mbox{$^*$Department of Physics and State Key Laboratory of Nuclear Physics
and Technology, }\\
\mbox{Peking University, Beijing, 100871, China}\\
\mbox{$^\dagger$High Energy Division, Argonne National Laboratory, Argonne, IL 60439, USA}\\
\mbox{$^\S$Department of Physics, National Taiwan Normal University, Taipei, 116, Taiwan}\\
\mbox{$^\$$Illinois Institute of Technology, Chicago, Illinois 60616-3793, USA}
}

\begin{abstract}
Forward-backward asymmetries $A_{FB}^t$ and $A_{FB}^\ell$ are
observed in the top quark $t$ rapidity distribution and in the
rapidity distribution of charged leptons $\ell$ from
top quark decay at the Tevatron proton-antiproton collider, and a charge 
asymmetry $A_C$ is seen in proton-proton collisions at the Large Hadron Collider 
(LHC). In this paper, we update our previous studies of the Tevatron asymmetries 
using the most recent data. We provide expectations for $A_C$ at the LHC based 
first on simple extrapolations from the Tevatron, and second based on 
new physics models that can explain the Tevatron asymmetries. We examine the 
relationship of the two asymmetries $A_{FB}^t$ and $A_{FB}^\ell$.  We show their
connection through the $(V-A)$ spin correlation between the charged lepton and
the top quark with different polarization states. We show that the ratio of the two 
asymmetries provides independent insight into the physics interpretation 
of the top quark asymmetry. We emphasize the value of the 
measurement of both asymmetries, and we conclude that a model which produces 
more right-handed than left-handed top quarks is suggested by the 
present Tevatron data.
\end{abstract}

\maketitle

\section{Introduction}
\label{sec: intro}

The observation of a larger than expected forward-backward asymmetry $A_{FB}
^t$  in the rapidity of top quarks produced at the Fermilab Tevatron collider~\cite
{Aaltonen:2011kc, Abazov:2011rq} continues to hold considerable attention in the 
community of particle physicists.  It is one of few manifestations of a deviation 
from predictions of the standard model (SM).  That the deviation occurs in the top 
sector suggests that its interpretation might well involve new physics (NP), given 
that the large mass of the top quark is comparable in value to the electroweak 
scale.   Indeed, many NP models have been proposed to explain the 
enhancement of $\afbt$.  These models usually postulate the existence of new 
states, whether in the direct-channel coupling to $t \bar{t}$, or exchanged in a 
cross-channel and coupling the top quark to first- and/or second-generation 
quarks.  Examples include flavor-changing $Z^{\prime}$~\cite{Jung:2009jz,Cao:2009uz,
Xiao:2010hm,Cao:2011ew,Berger:2011ua,Zhou:2011dg,AguilarSaavedra:2011vw,Gresham:2011pa,
AguilarSaavedra:2011zy,AguilarSaavedra:2011hz,Duraisamy:2011pt,AguilarSaavedra:2011ug,Gresham:2011fx,Ko:2011vd,Ko:2011di,Ko:2012gj,
Ko:2012ud,Li:2012dtz,Ko:2012mk,Alvarez:2012ca,Drobnak:2012rb,Aguilar-Saavedra:2013rza}, 
$W^{\prime}$~\cite{Cheung:2009ch,Barger:2010mw,Cheung:2011qa,Barger:2011ih,
Bhattacherjee:2011nr,Craig:2011an,Chen:2011mga,Cao:2011hr,Yan:2011tf,Knapen:2011hu,
Duffty:2012zz,Adelman:2012py,Endo:2012mi}
and axigluon $G^{\prime}$~\cite{Antunano:2007da,Ferrario:2008wm,Ferrario:2009bz,
Ferrario:2009ns,Frampton:2009rk,Cao:2010zb,Chivukula:2010fk,Xiao:2010ph,
Bai:2011ed,wang:2011taa,Tavares:2011zg,Alvarez:2011hi,AguilarSaavedra:2011ci,Wang:2011hc,
Krnjaic:2011ub,AguilarSaavedra:2011ck,Gabrielli:2011zw,Zhu:2012um,Cvetic:2012kv,
Dutta:2012ai,Gross:2012bz,Gresham:2012kv} models, 
among others ~\cite{Djouadi:2009nb,Shu:2009xf,Arhrib:2009hu,Dorsner:2009mq,Bauer:2010iq,
Chen:2010hm,Alvarez:2010js,Patel:2011eh,Zerwekh:2011wf,Barreto:2011au,Foot:2011xu,Jung:2011zv,
Jung:2011ua,Babu:2011yw,Djouadi:2011aj,Hektor:2011ms,Cui:2011xy,Frank:2011rb,Davoudiasl:2011tv,Jung:2011id,Kosnik:2011jr,delaPuente:2011iu,
Wang:2011mra,Endo:2012mi,Drobnak:2012cz}.

Strong constraints on models of new physics come from a variety of sources, 
whether from low-energy precision data that limit flavor changing couplings of the 
top quark, or from collider data such as the $t \bar{t}$ invariant mass distribution 
and the $t \bar{t}$ total cross section at the Tevatron.  Models of NP also 
face experimental constraints from searches for new phenomena at the LHC such 
as the absence of direct evidence thus far for new heavy gauge bosons $W'$ 
~\cite{Chatrchyan:2012su} and 
$Z'$, and strong bounds on the cross section at the LHC for the production of pairs 
of same-sign top quarks~\cite{Aad:2012bb,Chatrchyan:2012sa}.    

Of particular interest to us have been the implications of models of new physics for 
the polarization of the top quark, and methods that can be used to measure the 
polarization~\cite{Berger:2012an}.  
This focus on the top quark polarization also serves as a unifying 
theme for the topics discussed in this new paper.   In the SM, strong production of 
$t \bar{t}$ pairs in quantum chromodynamics (QCD) yields an equal number of 
positive and negative helicity top quarks, hereafter referred to as $t_R$ and $t_L$.  
Electroweak production in single top quark production, for example, yields 
primarily $t_L$.  Therefore, a demonstration that a significant fraction of top quarks 
are produced with positive helicity would herald new physics.   

In addition to $A_{FB}^t$ of the top quark, the D0 group reports a positive forward-backward asymmetry of charged 
leptons from top quark decays.   The measurement is done in two ways~\cite{Abazov:2011rq,:2012bfa}, both based 
on data corresponding to an integrated luminosity of $5.4\rm{~fb}^{-1}$.   The value $A_{FB}^\ell=(15.2\pm 4.0)\%$ 
is measured in the {$\ell$+jets} final states~\cite{Abazov:2011rq}.  The second method uses the dilepton final states 
from $t \bar{t}$ production, where the $W$ bosons from the $t$ and $\bar{t}$ decays both decay leptonically.  The 
result obtained is $A^{\ell}_{\rm FB}= (5.8 \pm 5.1 ({\rm stat}) \pm 1.3 ({\rm syst}))$\%.  A combination of the two 
measurements yields $A^{\ell}_{\rm FB}= {11.8 \pm 3.2}\%$.   The combined result may be compared with the values 
$(2.1\pm0.1)\%$ from simulations of the SM or $(4.7 \pm 0.1)\%$ once QCD+EW corrections 
are included~\cite{Bernreuther:2010ny,:2012bfa}, an excess at the level of 2.2 standard deviations. 
In a previous paper,  we investigated the kinematic and dynamic relationship 
between the two asymmetries 
$A_{FB}^t$ and $A_{FB}^\ell$~\cite{Berger:2012nw}.   The fact that $A_{FB}^\ell$ and 
$ A_{FB}^t$ are larger than the SM predictions indicates that the charged lepton 
strongly prefers to move in the same direction as the top quark from which it 
originates~\cite{Krohn:2011tw,Falkowski:2011zr}.  Data on the ratio of the two asymmetries 
tend to favor models in which more $t_R$ than $t_L$ are produced, but confirmation with greater 
statistical and systematic precision is desirable.  
A detailed analysis of the SM prediction of the lepton charge asymmetry
at the Tevatron and the LHC can be found in Ref.~\cite{Bernreuther:2012sx}.  

In this paper, we elaborate on the studies reported earlier and include new 
predictions.  We begin in Sec.~\ref{sec:update} with the definitions of the 
asymmetries measured at the Tevatron.  We summarize the Tevatron data and, 
using the latest data, we update our earlier fits in the framework of $Z'$, $W'$, and 
axigluon new physics models.   Unlike the Tevatron proton-antiproton collider, 
the LHC proton-proton collider offers no preferred direction for the measurement a rapidity 
asymmetry.   Nevertheless, a charge asymmetries $A_C^t$ for top quarks and 
$A_C^\ell$ for leptons can be defined and predicted in the SM.  
Using data from the Tevatron, we estimate what may be observed for these 
charge asymmetries at the LHC in 
the context of models of new physics, and we compare these expectations with 
LHC data in in Sec.~\ref{sec:lhc}.   As we show, despite limited statistics, the LHC 
data on the charge asymmetry are also consistent with a deviation from the SM, 
although perhaps not as great a deviation as expected from an extrapolation from 
the Tevatron observations.      

The relationship of $A_{FB}^t$ and $A_{FB}^\ell$ is addressed in 
Sec.~\ref{sec:landtasy} and in Appendix A where we include detailed derivations of 
results not published before.   
The essential starting point is the $V-A$ structure of the matrix element for the 
decay 
$t \rightarrow W^+ b \rightarrow b \ell^+ \nu$.   Section~\ref{subsec:kin} contains 
a discussion of the angular distribution of decay lepton $\ell^+$, first in the rest 
frame of the top quark and then after the top quark is boosted in rapidity and 
transverse momentum.  We pay particular attention to the positive/negative helicity 
state of 
the top quark because the final momentum and angular distributions of leptons in 
the laboratory frame, after the top quark 
is boosted, depends significantly on the top quark's polarization state.  In 
Sec.~\ref{sec:landtasy}, we derive the relationship of the lepton asymmetry $A_{FB}^\ell$ and 
the top quark asymmetry $A_{FB}^t$ separately for the left- and right-handed 
polarization states of the top quark.   

Different models of new physics produce top 
quarks with different proportions of left- and right-handed polarization.  For 
example, $Z'$ and $W'$ models produce predominantly right-handed top quarks, 
whereas the axigluon model generates unpolarized top quarks.  We use an 
axigluon model and a $W^{\prime}$ model in Sec.~\ref{sec:scan2} to deduce 
their different expectations for the ratio of the lepton and top quark asymmetries.  
In the case of both models, the allowed parameters produce a range of values for the 
ratios $A_C^\ell/ A_C^t$ at the LHC and $A_{FB}^\ell/A_{FB}^t$ at the Tevatron, aligned 
along approximately straight lines in plots of $A_C^\ell$ vs $A_C^t$ and of $A_{FB}^\ell$ 
vs $A_{FB}^t$.  Ideally, precise data would provide a definite point in the two dimensional 
plot and tightly constrain the parameter space.  

Our conclusions appear in Sec.~\ref{sec:con}.   We emphasize the value of 
making measurements both of $A_{FB}^t$ and $A_{FB}^\ell$ and of  
$A_C^\ell$ and $A_C^t$ because their ratio 
can be related through top quark polarization to the underlying dynamics of top 
quark production.   

\section{Tevatron Data and Updated Fits}\label{sec:update}

The top quark forward-backward asymmetry in $t\bar t$ pair production at the 
Tevatron is defined as
\be
A_{FB}^t = \frac{N(\Delta y>0) - N(\Delta y < 0)}{N(\Delta y>0) + N(\Delta y < 0)},
\label{eq:def_afbt}
\ee
where $\Delta y=y_t-y_{\bar t}$ is the difference between the rapidities of the top quark 
and the anti-top quark, and $N(\Delta y>0)$ ($N(\Delta y < 0 )$) is the number of 
events with $\Delta y>0$ ($\Delta y<0$).  The proton beam is chosen as the 
direction of positive $z$. 
In the SM, the asymmetry is induced by perturbative diagrams beyond the leading 
order.  It is predicted to be $(8.7\pm 1.0)\%$, including NLO EW and 
QCD corrections~\cite{Kuhn:1998jr,Kuhn:2011ri}. 
The most recent D0 result in the $t\bar t$ rest frame is $\afbt = (19.6\pm 6.5)\%$~
\cite{Abazov:2011rq}, based on their $5.4~\rm{fb}^{-1}$ luminosity data set, while 
the measurement from CDF  is $(16.2\pm 4.7)\%$ based on their data set with 
integrated luminosity of $8.7~\rm{fb}^{-1}$~\cite{Aaltonen:2012afb}. 
CDF also reports that $\afbt$ in the region of large $t\bar t$ pair invariant mass 
($m_{t\bar t} \geq 450~ \rm{GeV}$) exceeds the SM prediction ($\sim 3\sigma$), 
although the significance is not as large as the $3.4 \sigma$ deviation of CDF's 
previous result~\cite{Aaltonen:2011kc}.
More explicitly,  $\afbt (m_{t\bar t} \geq 450~ \rm{GeV}) = (29.6\pm6.7)\%$ and the 
SM prediction  is $(12.8\pm1.1)\%$~\cite{Kuhn:2011ri}.

Many new physics models have been proposed to explain the discrepancy of 
$\afbt$ between data and the SM prediction.  Some of these models 
are now quite sophisticated.  It is not our intention in this paper to investigate models 
in detail.  Rather, we explore a few simple models as illustrations of a range of 
possibilities.  We begin in this section with an update of our previous fits 
to Tevatron data for three models: flavor-changing $Z'$ exchange, flavor-changing 
$W'$ exchange, and axigluon models.   The minimal version of the $Z'$ model 
implies a large rate for same-sign top quark pair production at the LHC, not 
supported by data~\cite{Berger:2011ua,Aad:2012bb,Chatrchyan:2012sa}.  The $W'$ model is 
highly constrained by data on the $t\bar t$ plus jets final state at the 
LHC~\cite{Knapen:2011hu,Duffty:2012zz,Chatrchyan:2012su}.

The effective interaction between a flavor-changing $Z^\prime$/$W^\prime$ and SM particles is 
\be
\mathcal{L}=\frac{g_2 f_{V^\prime}}{\sqrt{2}} \bar{q}\gamma^\mu P_R t V^\prime_{\mu} + h.c.~, 
\label{eq:wzprime}
\ee
where $g_2$ is the weak coupling, and $q=u~(d)$ for $V^\prime = Z^\prime (W^
\prime)$.
In addition to the SM process $q\bar{q} \to g \to t\bar{t}$ and its NLO corrections, 
the $t\bar{t}$ pair will also be produced via a $t$-channel process with
a $Z^\prime$ or $W^\prime$ mediator.  Using   
``$\lambda_i=+$" to represent the positive helicity of particle $i$ (right-handed 
polarization for massless particle), and ``$\lambda_i=-$" 
the negative helicity (left-handed polarization for massless particle),  we express 
the helicity amplitude 
$M_{V^\prime}^{t}(\lambda_q,\lambda_{\bar q}, \lambda_{t},\lambda_{\bar t})$,  
apart from a common factor 
$\displaystyle{\frac{g_2^2f_{V^\prime}^2\hat{s}\delta_{ac}\delta_{bd}}{8\left(\hat{t}-m_{V^\prime}^2\right)}}$, 
where $a,~b,~c$ and $d$ are the color indexes of $q,~\bar{q},~t$ and $\bar{t}$, as
\bea
\mathcal{M}_{V^\prime}^t(+---)&=&\left(2+r_V^2\right)\sqrt{1-\beta^2} \sin\theta, \nonumber \\
\mathcal{M}_{V^\prime}^t(+--+)&=&\left[2(1-\beta)+r_V^2(1+\beta)\right](1-\cos\theta),\nonumber \\
\mathcal{M}_{V^\prime}^t(+-+-)&=&-\left[2(1+\beta)+r_V^2(1-\beta)\right](1+\cos\theta), \nonumber \\
\mathcal{M}_{V^\prime}^t(+-++)&=&-\left(2+r_V^2\right)\sqrt{1-\beta^2} \sin\theta~.
\label{eq:wzp_hel}
\eea 
The variables $\hat s$ and $\hat t$ are the usual Mandelstam variables, $m_{V'}$ is the mass of $V^\prime$,
 $r_V=m_t/m_{V^\prime}$, $\beta=\sqrt{1-4m_t^2/\hat{s}}$, and $\theta$ is 
the polar angle of the top quark in the center mass (c.m. frame) 
of the $t\bar{t}$ pair, measured relative to the initial state quark. 
In the highly boosted limit of $\beta\to 1$, the nonzero helicity amplitudes are
\bea
\mathcal{M}_{V^\prime}^t(+--+)&\sim&2r_V^2(1-\cos\theta), \nonumber \\
\mathcal{M}_{V^\prime}^t(+-+-)&\sim&-4(1+\cos\theta)~.
\label{eq:wzp_hel2}
\eea

For the axigluon ($G^\prime$) model, we assume, for simplicity, that the interaction of 
the axigluon with the SM quarks is purely pseudo-vector-like and can be written as
\be
\mathcal{L} = g_s \left(g_l~\bar{q_i}\gamma^\mu \gamma_5 q_j
+ g_h~\bar{Q_i}\gamma^\mu \gamma_5 Q_j\right)t^A_{ij} G^{\prime~A}_{\mu},
\label{eq:gprime}
\ee
where $t_{ij}^A$ is the generator of the color $SU(3)$ group; $q$ denotes the first 
two generation quarks in the SM and $Q$ the third
generation quarks.   The coupling $g_s$ is the usual strong coupling strength; 
$g_l$ and $g_h$ are the 
coupling parameters of the axigluon to 
the light quark ($q$, i.e. first two generations) and the heavy quark ($Q$, i.e. third 
generation), respectively.   

The process $q\bar{q}\to G^\prime \to t\bar{t}$ contributes to $t\bar t$ production at 
hadron colliders.  Its helicity amplitudes  
$M_{G^\prime}(\lambda_q,\lambda_{\bar q},\lambda_t,\lambda_{\bar t})$ are
\bea
\mathcal{M}_{G^\prime}(+-+-) &=& \mathcal{M}_{G^\prime}(-+-+)= \frac{g_s^2t^A_
{ba}t^A_{cd}\hat{s}~(-g_lg_h)}{\hat{s}-m_{G^\prime}^2 +i m_{G^\prime} \Gamma_
{G^\prime}} \beta (1+\cos\theta),\nonumber\\
\mathcal{M}_{G^\prime}(-++-) &=& \mathcal{M}_{G^\prime}(+--+)=\frac{g_s^2t^A_
{ba}t^A_{cd}\hat{s}~(-g_lg_h)}{\hat{s}-m_{G^\prime}^2 +i m_{G^\prime} \Gamma_
{G^\prime}} \beta (1-\cos\theta).
\label{eq:gp_hel}
\eea
where $\Gamma_{G^\prime}$ is the width of axigluon. For $m_{G^\prime} > 
2m_t$, which is the case in our study, 
\begin{equation}
\Gamma_{G^\prime}=\frac{\alpha_Sm_{G^\prime}}{6}\left\{4g_l^2+g_h^2\left[1+
\left(1-\frac{4m_t^2}{m_{G^\prime}^2}\right)^{3/2}\right]\right\}.
\end{equation}
For coupling strength $g_l = g_h = 1$, the ratio $\Gamma_{G^\prime}/m_{G^\prime} \sim 0.1$.  

The absence of pronounced deviations from the SM expectation in the measured
$m_{t\bar{t}}$ distribution~\cite{Aaltonen:2011kc, Abazov:2011rq}
indicates that the axigluon should be heavy and/or broad.   Since the term linear in 
$\cos\theta$ appears only in the interference term, the contribution to $\afbt$ in $t
\bar{t}$ production from an axigluon is therefore 
through interference with the SM channel.  Its effect becomes important  in the 
region of large $m_{t\bar{t}}$,  
i.e. $\beta \sim 1$.  
The interference term in the overall squared amplitude is proportional to 
\be
\left[2g_lg_h(1+\cos\theta)^2-2g_lg_h(1-\cos\theta)^2\right]\frac{\hat{s}(\hat{s}-m_
{G^\prime}^2)}{(\hat{s}-m_{G^\prime}^2)^2+m_{G^\prime}^2\Gamma_{G^\prime}
^2} = \frac{4g_lg_h\hat{s}(\hat{s}-m_{G^\prime}^2)\cos\theta}{(\hat{s}-m_{G^
\prime}^2)^2+m_{G^\prime}^2\Gamma_{G^\prime}^2}.
\ee
When an axigluon is heavy such that $\sqrt{\hat{s}}<m_{G^\prime}$, the product of 
$g_l g_h$ must be negative to obtain a positive 
$A_{FB}$~\cite{Ferrario:2009bz,Frampton:2009rk,Cao:2010zb}.  

We fit data at the Tevatron to determine the parameters of the three new physics 
models under consideration.   The SM contributions at NLO are included along 
with the contributions from the new physics models.  
We choose to fit the measured inclusive total cross section for $t\bar t$ production 
$7.5\pm0.31({\text{stat}})\pm0.34({\text{syst}})\pm0.15({\text{Z~theory}})
~{\rm pb}$~\cite{Aaltonen:2009tt}, and $\afbt$ from CDF 
with $8.7~\rm{fb}^{-1}$ integrated luminosity~\cite{Aaltonen:2012afb}.  We scan 
the parameter 
space of the models requiring that the predictions fit the total cross section 
as well as $\afbt$ for both $m_{t \bar t} <~450~{\rm GeV}~(7.8\%\pm 5.4\%)$ and 
$m_{t\bar t} \geq~450~{\rm GeV}~(29.6\%\pm 6.7\%)$ within $2\sigma$ accuracy.  
The SM  $t\bar t$ cross section we adopt is 
$6.77\pm 0.73~ \rm{pb}$ calculated with MCFM6.2~\cite{mcfm}.  For the SM 
predictions of $\afbt$ in different 
energy bins, we follow the values shown in Ref.~\cite{Kuhn:2011ri}: 
\be
A_{FB}^t (m_{t\bar t} <~450~{\rm GeV}) = (6.2\pm0.4)~\%,~~A_{FB}^t (m_{t\bar t} \geq~450~{\rm GeV}) = (12.8\pm1.1)~\%.
\label{eq:sm_afbt}
\ee
In addition to the observables listed above, there are also differential cross sections in the 
invariant mass $m_{t \bar{t}}$ and in the transverse momentum $p_{T_{t \bar{t}}}$ of the $t \bar{t}$ system.  Including such 
data in our fits would arguably provide further constraints on the allowed parameters of the models.   On the other hand, 
contributions from new physics tend to  affect the mass distribution at large values of $m_{t \bar{t}}$, where statistics are 
relatively poor and therefore less constraining on fits.   Moreover, and more importantly, to do a proper analysis, one 
would want to compute the new physics contributions at NLO, include the effects of parton showering, and model the 
experimental acceptance cuts whose effects are particularly significant at larger values of $m_{t \bar{t}}$.  A complete 
analysis in terms of new physics models is also complicated by the fact that data on the $m_{t \bar{t}}$ distribution are 
unfolded in terms of the SM shape and cut efficiencies.  
Even without extending our study to include data on differential cross sections, we find significant constraints on the 
coupling strengths of the models, as summarized below for the Tevatron and in the next section for the LHC.  Data from 
the LHC on same-sign top quark production, on the production of a pair of top quarks plus one jet, and on searches for 
dijet resonances are used to limit the space of allowed parameters.  The distinct features of the constrained models are 
instructive, as we show for the correlation between the two asymmetries $A_{FB}^t$ and $A_{FB}^\ell$, a reflection of 
their polarization predictions.
\begin{figure}[!htb]
\includegraphics[scale=0.26]{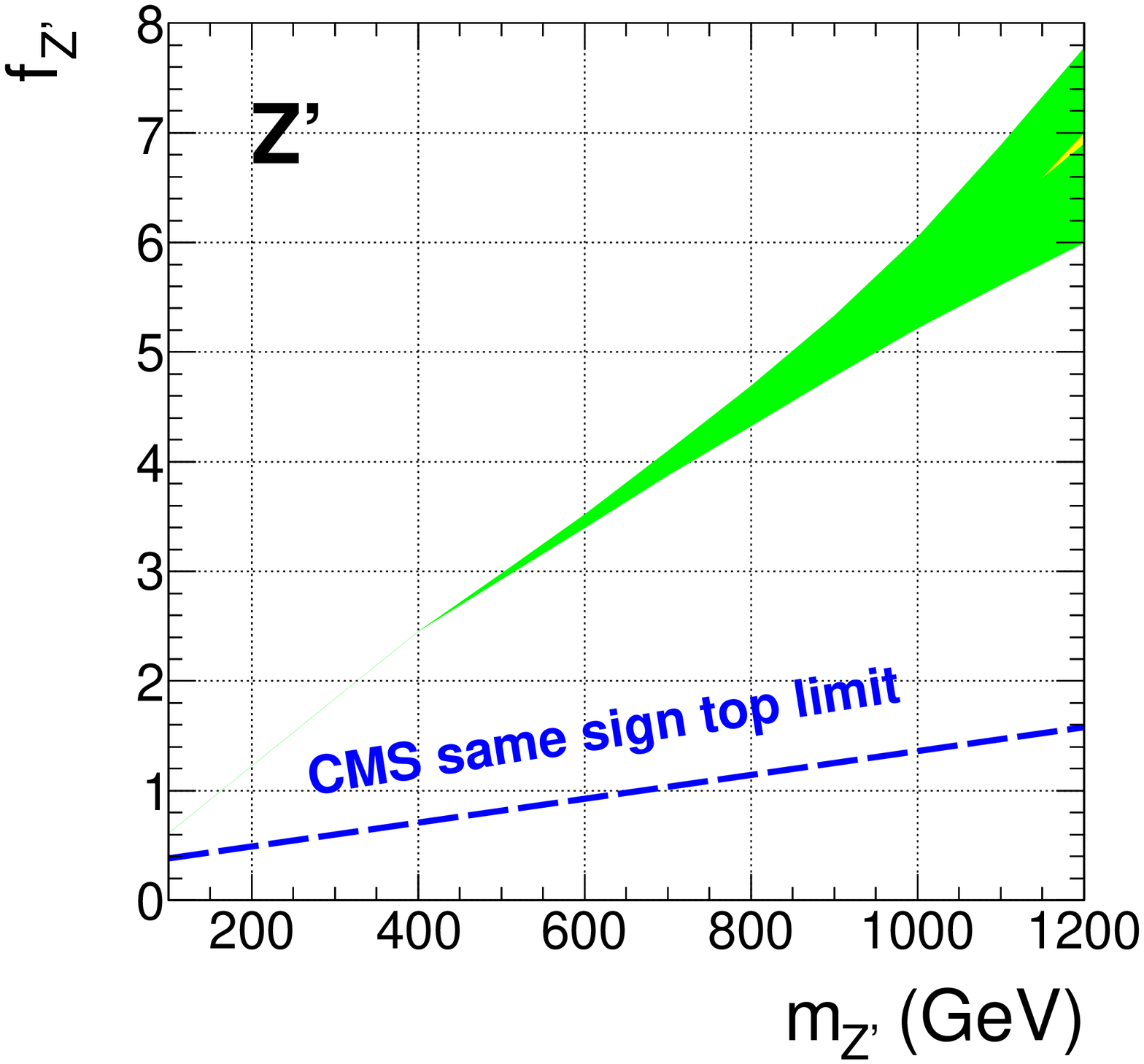}
\includegraphics[scale=0.26]{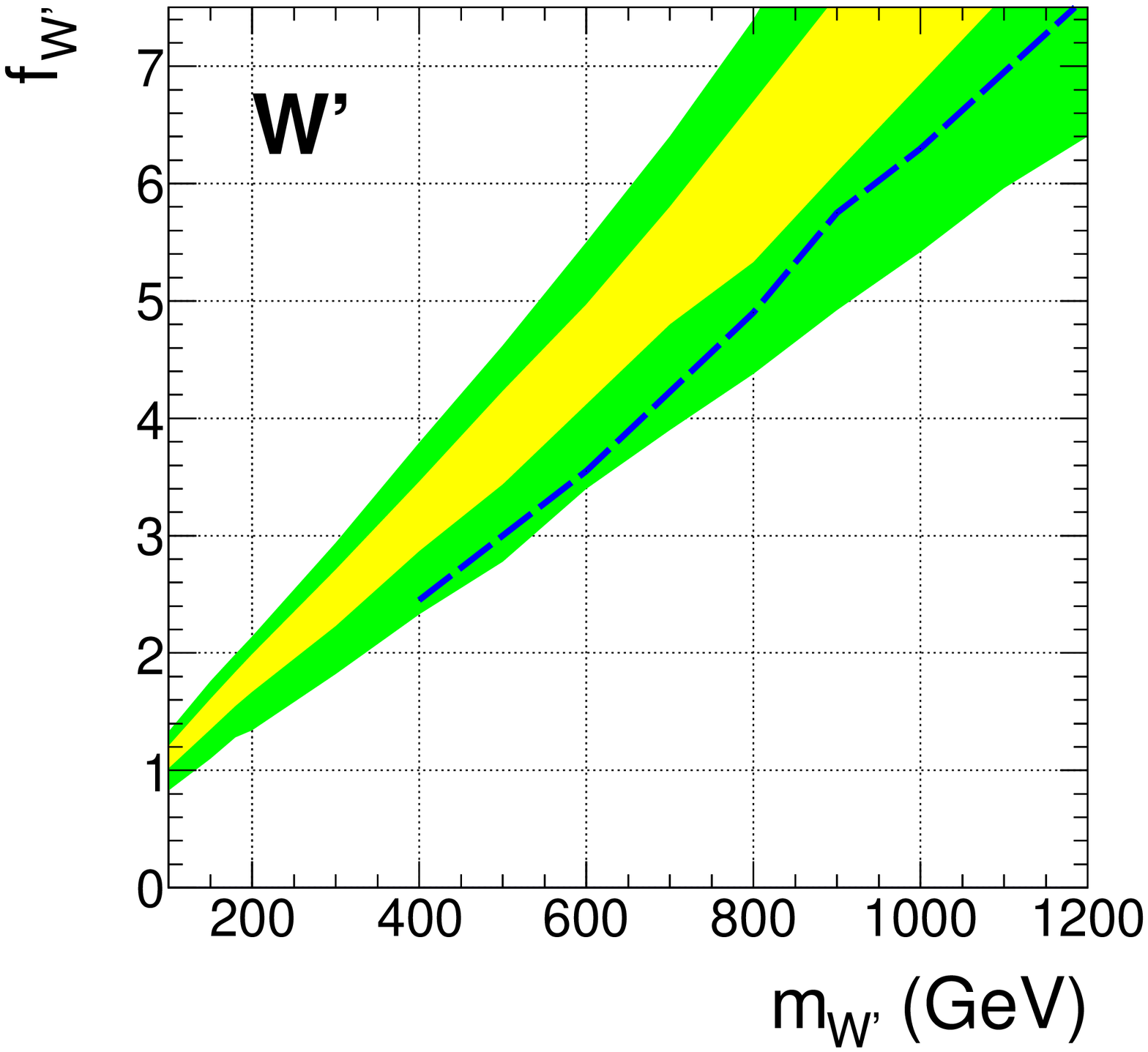}
\includegraphics[scale=0.26]{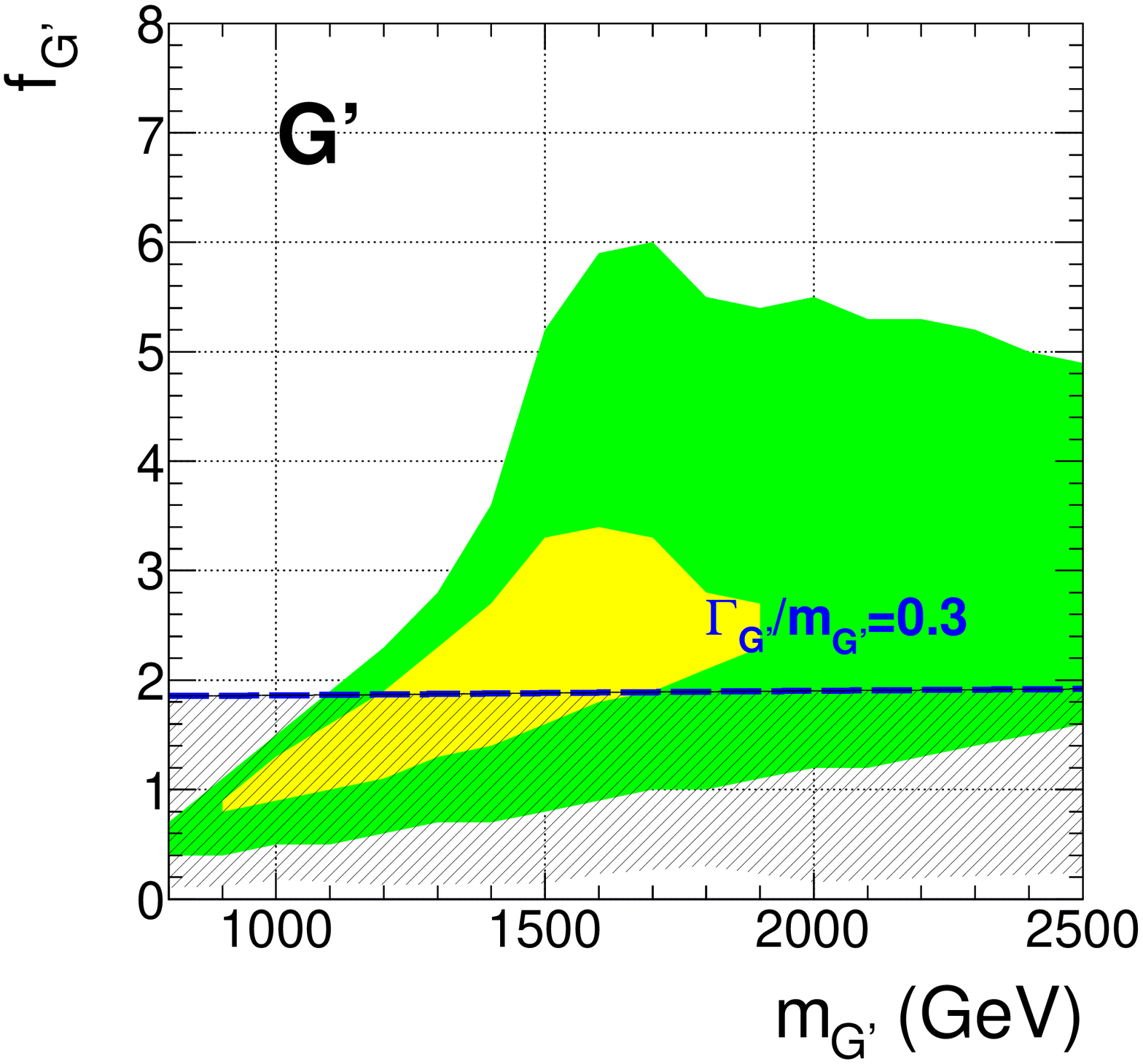}
\caption{The parameter space of three new physics models determined from fits to 
the Tevatron $t\bar t$ total cross section and $\afbt$ measured by the CDF 
collaboration in the intervals $m_{t\bar t} <~450~{\rm GeV}$ and $m_{t\bar t} 
\geq~450~{\rm GeV}$.  The inner shaded (yellow) region fits the data within $1\sigma$ 
and the outer shaded (green)  
region fits within $2\sigma$: flavor-changing $Z^\prime$ model, flavor-changing 
$W^\prime$ model, and axigluon model. The dashed line in (a) shows the bound 
obtained from the CMS search for same-sign top quark pairs at the LHC.   The dashed 
line in (b) shows the bound on the coupling $f_{W'}$ that we obtained from an analysis 
of the CMS data on top-pair-plus-one-jet events at the LHC.  The cross-hatched (blue) 
region in (c) is inferred from the limits set by ATLAS on axigluons from the search 
for enhancements in the dijet mass distribution.}
\label{fig:tevfit}
\end{figure}

Figure~\ref{fig:tevfit} shows the results of our fits for the three models. 
We simulate the models using MadGraph5~\cite{Alwall:2011uj}. 
The yellow (green) band is the parameter space which fits the Tevatron $t\bar t$ total cross 
section and $\afbt$ within $1(2)\sigma$.  For the $Z^\prime$ model, there is no 
allowed parameter space when $m_{Z^\prime}$ is less than about $400~{\rm 
GeV}$, and only a tiny region can fit within $1\sigma$ when $m_{Z^\prime}$ is 
heavier than $1100~{\rm GeV}$, with large couplings $f_{Z'}\gtrsim 6.6$ .  
This conclusion differs from the one in our earlier work.   The difference comes entirely 
from the fact that we are now fitting the most recent CDF data in which $\afbt$ is 
smaller.  The contribution from $Z^\prime$ exchange is fed by the $u$ and $\bar u
$ initial states which have large parton densities at the Tevatron $p\bar p $ collider. 
Therefore, both $\afbt$ and the $t\bar t$ total cross section change rapidly with the 
coupling $f_{Z^\prime}$.  When $f_{Z^\prime}$ is large enough to bring the total 
cross section into a region that is consistent with data at the $2\sigma$ level, $
\afbt$ at high $m_{t\bar t}~(\geq 450~{\rm GeV})$ becomes too large to fit the 
data.  As a result, only a small parameter space yields a better fit than the SM 
itself, and it is very difficult to reach agreement with data within $1\sigma$.  
Therefore, we now conclude that the minimal flavor-changing $Z'$ model can 
barely explain the large deviation of $\afbt$ from the SM observed at the Tevatron.   
In FIG.~\ref{fig:tevfit}($Z^\prime$), 
we also plot the upper limit of the coupling for the $Z^\prime$ model obtained from 
the search of same-sign top quark pairs at the 
LHC~\cite{Chatrchyan:2012sa}.  The region above the blue dashed curve is not 
allowed since too many same-sign top quark pairs would be produced.  The 
minimal version of the $Z^\prime$ model is definitely disfavored. 

In contrast to the $Z'$ case, there is a large region of parameter space in which 
the $W'$ model can fit the Tevatron data  within $1\sigma$ and $2\sigma$, as 
shown in  FIG.~\ref{fig:tevfit}($W^\prime$).  We scan the coupling $f_
{W'}$ in the $W'$ model up to 7.5 in our numerical study.\footnote{The upper 
bound on the coupling is set here by the choice of a perturbative bound 
$g_2^2f_{W^\prime}^2/(2\times4\pi)\leq1$, which 
means $f_{W^\prime}\leq7.7$~.} We see that the $W'$ model can fit data quite well 
with $m_{W'} \lesssim 1~{\rm TeV}$ for the coupling parameter 
$f_{W'} \lesssim 7.5$.  The 
asymmetry and the $t\bar t $ cross section do not change as sharply with coupling 
in the $W'$ model as they do in the $Z'$ case since their contributions are fed by the 
smaller $d$ and $\bar d$ parton densities. 
The upper limit of the coupling for the $W^\prime$ model is shown in the figure.   
We obtain this upper limit from an analysis of the CMS data on the 
production of a top pair plus one jet ~\cite{Chatrchyan:2012su}.  The region above 
the blue dashed curve is not allowed since too many $t\bar t+j$ events would be 
produced.  The data were not analyzed for values of the $W^\prime$ mass below 
400~GeV so we do not show a constraint below this value. 
There is a similar constraint from 
ATLAS~\cite{PhysRevD.86.091103}, but we do not use the bound shown in their 
FIG. 6 because the interference term between the SM and the $W^\prime$ model 
is not considered in the determination of their bound.  This interference is not 
negligible~\cite{Duffty:2012zz,Endo:2012mi}. We see that some region of the 
parameter space of the $W^\prime$ model remains open. 
The contribution to $\afbt$ at the Tevatron from top-$W^\prime$ associated 
production is not incorporated in our study since it is small at Tevatron energies owing to 
phase space and gluon parton distribution function suppression.

In the axigluon case, we scan $g_{l}$ and $g_{h}$ up to $8$.  For simplicity, 
we fix $g_l = -g_h=f_{G^\prime}$ in Eqs.(6-7). 
To achieve good agreement with data at the $1\sigma$ level, the mass of axigluon 
is required to be in the range of about $900~{\rm GeV}$ to $1900~{\rm GeV}$.  
For other axigluon masses, the model can only fit data at the $2\sigma$ level.  
These results are shown in  FIG.~\ref{fig:tevfit}($G^\prime$).   
In FIG.~\ref{fig:tevfit}($G^\prime$), we 
also show some bounds on axigluon masses and couplings obtained from a 
search for resonances in the dijet invariant mass distribution~
\cite{Aad:2011aj,Aad:2011fq,ATLAS:2012joa,ATLAS:2012vua}.   To obtain 
the lower bound on the coupling constant $f_{G^\prime}$, we generate parton level
dijet events in the axigluon model using MadGraph5 and MadEvent \cite{Alwall:2011uj}.   
After adding the cuts on the final state partons employed 
in~\cite{Aad:2011aj,Aad:2011fq,ATLAS:2012joa,ATLAS:2012vua}, we obtain 
the cross sections $\sigma\times\mathcal{A}$, where $\mathcal{A}$ represents acceptance.   
Comparing these results with 
the exclusion bound in~\cite{Aad:2011aj,Aad:2011fq,ATLAS:2012joa,ATLAS:2012vua}, 
we derive the lower bounds of the excluded region for $f_{G^\prime}$ as a function of 
axigluon mass, shown in FIG~\ref{fig:tevfit}($G^\prime$).
On the other hand, axigluons with large width cannot be 
excluded using the search technique described in the ATLAS paper.  The 
contribution from a broad axigluon 
would cover a large fraction of the search region in the dijet invariant mass and be 
absorbed into the data-driven background fit.   To 
account for this limitation of the search, we sketch a soft upper limit of the 
exclusion region in $f_{G^\prime}$ determined by the value 
$\Gamma_{G^\prime}/m_{G^\prime} = 0.3$ (the blue shaded region in FIG.~\ref
{fig:tevfit}($G^\prime$)).

Before concluding this discussion of fits to the Tevatron data, we acknowledge limitations 
of our approach.  For the three new physics models, we compare the $t\bar t$ forward-backward 
asymmetry with the unfolded data of the CDF collaboration.  The unfolded result is obtained under 
the assumption that the events follow the SM $t\bar t$ event distribution, so the comparison is 
not exact for new physics models.  The correction could be significant for $t-$channel exotic vector 
bosons.   Indeed, the authors of Ref.~\cite{Gresham:2011pa} show that the cut efficiency is larger 
in the SM in the region of large $m_{t\bar t}$ than for the case of a $t-$channel exotic vector boson.  
There are two main influences of this difference~\cite{Gresham:2011pa}.  First, the lower 
efficiency of the $t-$channel new physics models, especially in the large $m_{t\bar t}$ region, 
will suppress the number of large $m_{t\bar t}$ events in the new physics models and release 
the tension between theory and data.  However, in our analysis, we do not fit the differential 
cross section in $m_{t\bar t}$, only the cross section integrated over $m_{t\bar t}$.  
Since the cross section falls rapidly with $m_{t\bar t}$, the pertinent correction is relatively small
in our fit.   Second, the difference between the cut efficiencies for events with $\Delta y>0$ and 
$\Delta y<0$ in the new physics models will decrease the prediction of $A_{FB}^t$.  Such effects 
are shown in~\cite{Gresham:2011pa} to be not as large as the cut efficiency effect on the invariant 
mass distribution.   It is worth remarking that the NLO QCD correction for the pure new physics 
term and for the NP-SM interference term is larger in the large invariant mass region than in 
the low invariant mass region~\cite{Yan:2011tf}.   The NLO QCD correction will therefore 
counteract the cut efficiency effect at least partly.  A complete investigation that includes both 
the NLO and cut efficiency effects is desirable, but we judge that the simpler approach used here 
suffices for our limited purposes.

To summarize this section, we remark that based on the latest $\afbt$ data from 
CDF at the Tevatron, the simple $Z'$ model is disfavored, and a light $W'$ ($\lesssim 1~{\rm TeV}$) 
is preferred for a small coupling strength, while an axigluon model can give a good fit with 
an axigluon mass about $1200~{\rm GeV}\sim 1900~{\rm GeV}$.

\section{LHC proton-proton Collider}\label{sec:lhc}

In this section we address the charge asymmetry in rapidity $A_C$ measured at 
the LHC.  We obtain estimates of LHC expectations 
first by simple extrapolation from the Tevatron data on $\afbt$ and second based 
on the new physics models whose parameters we determine in 
Sec.~\ref{sec:update}.  

The proton-proton LHC collider is symmetric in rapidity, and it is ambiguous to 
define the forward or backward region.  However, the $u$ and  $d$ (valence 
quarks inside the proton) parton densities carry, on average, a larger fraction of 
the momentum of the proton than the $u$ and $d$ antiquark densities (sea quarks 
inside the proton).  With the knowledge that there is a forward-backward 
asymmetry in the 
perturbative production process for $q\bar{q} \rightarrow t\bar t$ production, we 
expect that the top quark at the LHC will be boosted in the direction of the incident 
quark.  As a result, top quarks should accumulate in the region of large rapidity 
and anti-top quarks will be preferentially in the central region. Therefore, one can 
define an asymmetry $A_C$ at the LHC as
\be
A_{C} = \frac{N(|y_t|>|y_{\bar t}|) - N(|y_t|<|y_{\bar t}|)}{N(|y_t|>|y_{\bar t}|) + N(|y_t|<|y_{\bar t}|)}.
\label{eq:def_afbtlhc}
\ee
The SM prediction including NLO EW and QCD contributions is $A_C = 0.0115$ at 
$7$ TeV center-of-mass energy~\cite{Kuhn:2011ri}, and the predicted value drops 
when the collider energy increases.  The event generator MC@NLO provides a 
slightly different result, $A_C = 0.006$~\cite{ATLAS-CONF-2012-057}, owing to 
different normalization and the 
absence of NLO EW corrections.

Recent measurements of $A_C$ at the LHC have been published by the CMS and 
ATLAS collaborations based on data sets with 
$4.7~{\rm fb}^{-1}$ of integrated luminosity.  The results from CMS~\cite{CMS-PAS-TOP-11-030} 
obtained from the lepton plus jet final state and 
ATLAS ~\cite{ATLAS-CONF-2012-057} obtained from combining both lepton plus jet and dilepton 
channels are  
\bea
{\rm CMS~(lepton+jets):}~ 0.004\pm0.010{\rm (stat.)}\pm 0.012{\rm (syst.)}\nonumber\\
{\rm ATLAS~(combined):}~ 0.029\pm0.018{\rm (stat.)}\pm 0.014{\rm (syst.)}.
\label{eq:lhc_afbt}
\eea
The ATLAS central value is an order of magnitude larger than the CMS value, but they 
agree within the large uncertainties in both experiments, and they are consistent with the SM prediction.  

At the LHC, $t\bar t$ production is dominated by the gluon-gluon initial state which 
provides no asymmetry, and the asymmetry $A_C$ 
generated by the quark-antiquark initial state is therefore expected to be diluted 
substantially.   An approximate estimate for the LHC asymmetry is  
\bea
A_C^t &\approx& \frac{\sigma(q\bar{q}\to t\bar{t})}{\sigma(gg\to t\bar{t})+\sigma(q\bar{q}\to 
t\bar{t})}\times A_{FB}^{t}(q\bar{q}\to t\bar{t}) \times \tilde\varepsilon. 
\label{eq:ac_approx}
\eea
The first term represents the fraction of the top-quark pair production cross section 
induced by the $q\bar{q}$ initial state which is about 17~\% in the SM at 7~TeV 
LHC.  The second term is the asymmetry induced by the $q\bar{q}$ initial state. 
Given that about $88\%$ of the $t\bar t$ production cross section in the SM 
comes from the $q\bar q$ initial state at the Tevatron,  $A_{FB}^{t}(q\bar{q}\to t\bar
{t})$ can be extracted from the top quark forward-backward asymmetry observed 
at the Tevatron; we use $A_{FB}^{t}(q\bar{q}\to t\bar{t})\approx \afbt/88\%$, where 
$\afbt$ is the measured top quark asymmetry. 
The last term $\tilde\varepsilon$ in Eq.~(\ref{eq:ac_approx}) represents the probability 
of correct identification of  the forward direction, 
namely how frequently the forward direction represents the direction of the initial 
state quark.  This probability has to be evaluated for both the Tevatron and the 
LHC.  

At the Tevatron, the momentum of the proton beam is chosen as the forward 
direction.   Therefore, the probability is
\begin{equation}
\varepsilon^{TEV}=\frac{\sigma(q\bar q\to t\bar t)_{q/P,\bar q /\bar P}}{\sigma(q\bar q\to t\bar t)_{\rm total}},
\end{equation}
where the denominator is the total cross section of $q\bar q\to t \bar t$ and the 
numerator is the contribution to the total cross section when the initial state quark 
and antiquark come from proton and anitproton, respectively.   An explicit 
evaluation  can be obtained from the integral over parton densities: 
\begin{eqnarray}
\varepsilon^{TEV}(\hat{s}) =
\frac{\sum_q\int_{\hat{s}/S}^1f_{q/p}(x_1)f_{\bar q/\bar p}\left(\frac{\hat{s}}
{x_1S}\right)\frac{dx_1}{x_1}}{\sum_q\int_{\hat{s}/S}^1\left[f_{q/p}(x_1)f_{\bar q/\bar p}\left(\frac{\hat{s}}
{x_1S}\right)+f_{\bar q/p}(x_1)f_{q/\bar p}\left(\frac{\hat{s}}
{x_1S}\right)\right]\frac{dx_1}{x_1}} ,
\end{eqnarray}
where $S$ is the square of the total energy of the $\bar{p} p$ collision and 
$\hat{s}$ denotes the square of the energy in the partonic collison. 

At the LHC, with no preferred direction in a proton-proton collider, the boost 
direction of the $t\bar t$ system is chosen to be the forward direction. Hence, the 
probability of choosing the forward direction correctly is 
\begin{equation}
\varepsilon^{LHC}=\frac{\sigma(q\bar q\to t\bar t)_{q > \bar q }}{\sigma(q\bar q\to t\bar t)_{\rm total}},
\end{equation}
where the numerator now is the contribution to the total cross section when the 
initial quark momentum is larger the initial state antiquark momentum.   The 
corresponding integral over parton densities is
\begin{eqnarray}
\varepsilon^{LHC}(\hat{s}) = 
\frac{\sum_q\int_{\sqrt{\hat{s}/S}}^1f_{q/p}(x_1)f_{\bar q/p}\left(\frac{\hat{s}}
{x_1S}\right)\frac{dx_1}{x_1}}{\sum_q\int_{\hat{s}/S}^1f_{q/p}(x_1)f_{\bar q/p}\left(\frac
{\hat{s}}{x_1S}\right)\frac{dx_1}{x_1}}.
\end{eqnarray}
 
\begin{figure}[!htb]
\includegraphics[scale=0.26]{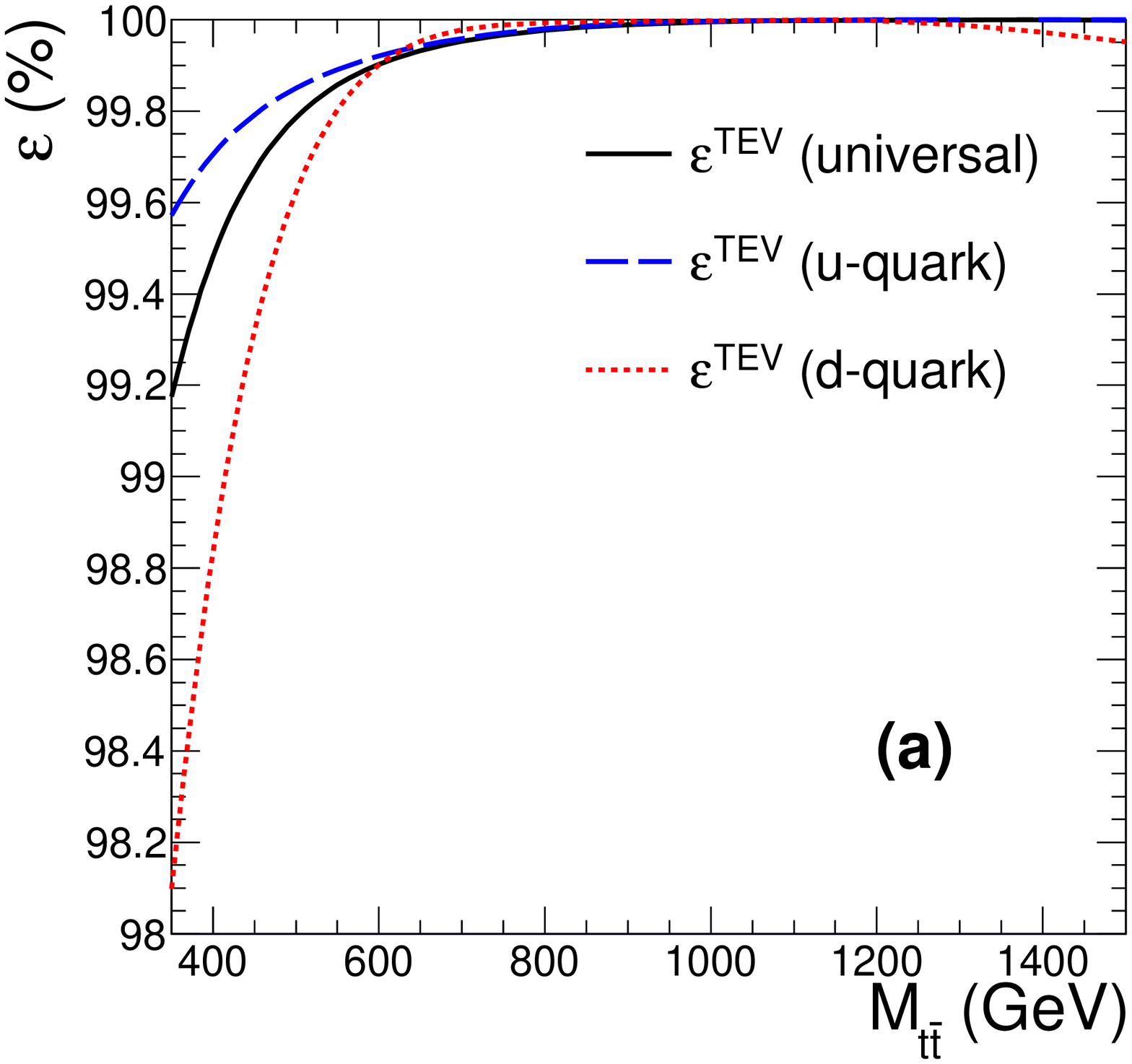}
\includegraphics[scale=0.26]{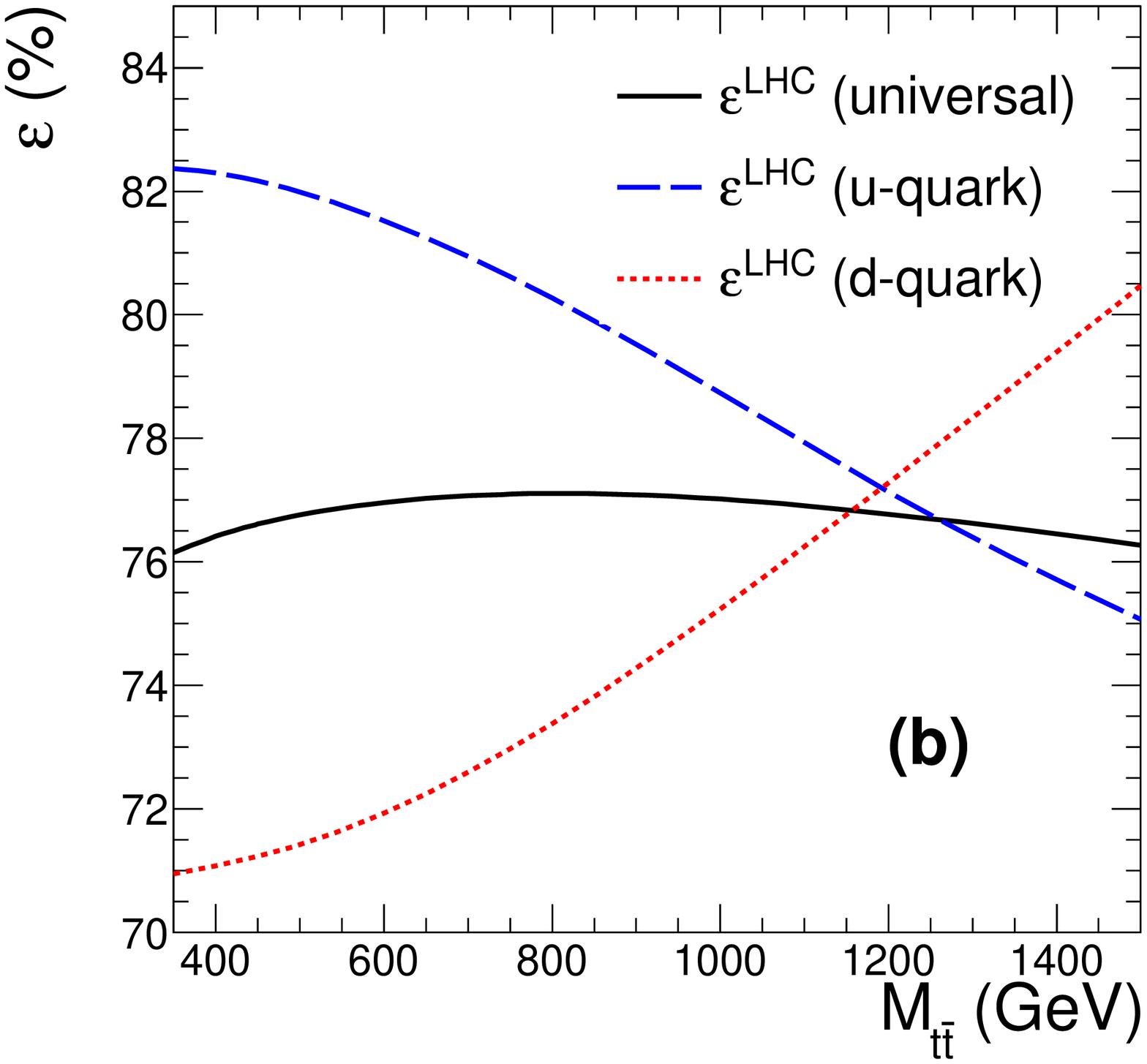}
\includegraphics[scale=0.26]{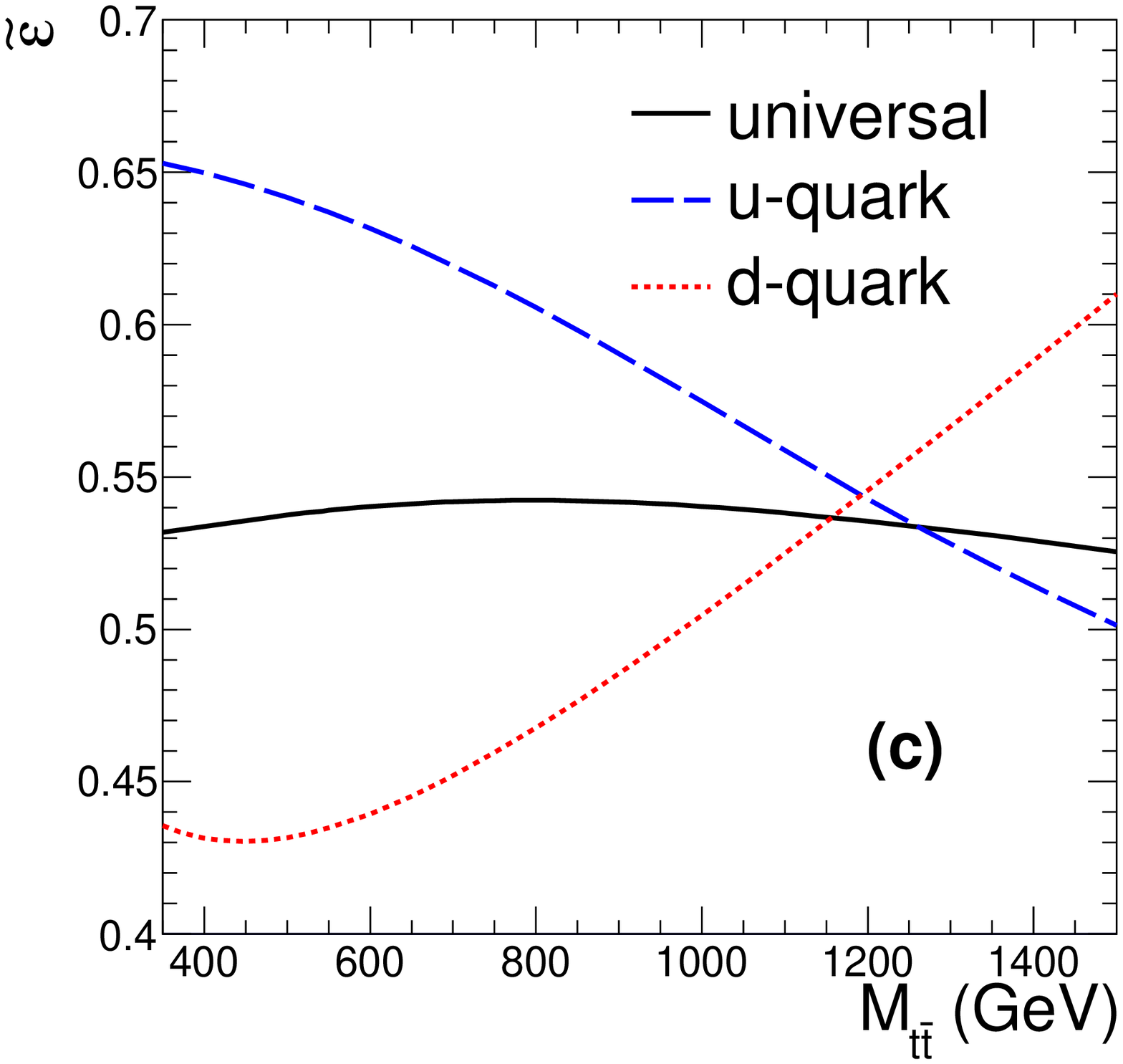}
\caption{(a) The probability for choosing the correct forward 
direction, the direction of initial state quark at the Tevatron. (b) The probability 
for choosing the correct forward 
direction, the direction of initial state quark at the LHC.   For 
the ``universal'' case, we include all four flavors of light quarks with equal  
weight when we calculate $\varepsilon$.  For the ``$u(d)$-quark'' case, we  
include only the contribution from the $u(d)$-quark.  We do not show the behavior of 
other light flavors because they are sea quarks at the Tevatron and do not give 
a significant contribution to $\afbt$.  (c) The relative suppression factor
$\tilde\varepsilon$ for the universal, $u$-quark, and $d$-quark cases.}
\label{fig:epsilon}
\end{figure}

We evaluate the efficiencies explicitly using the MSTW parton distribution 
 functions (PDFs)~\cite{Martin:2009iq}.  The efficiencies vary with the invariant mass of the $t\bar t$ system, as shown 
in Fig.~\ref{fig:epsilon}.   At the Tevatron, the value of $\varepsilon^{TEV}$ is nearly $100 \%$, 
and the proton (antiproton) beam represents the direction of initial quark (antiquark) quite well.  
However, at the LHC, the probability $\varepsilon^{LHC}$ that the initial quark direction 
matches the boost direction of the $t\bar t$ system is 
lower.   We find values in the range $72\%-83\%$, depending upon the initial state quark and the 
effective energy of the $t\bar t$ center mass system (FIG. ~\ref{fig:epsilon}).  Since the values of 
$\varepsilon$'s are not $100\%$ at the LHC,  the wrong choice of forward direction decreases the 
absolute value of $A_{FB}^t$. 

The measured number of forward (backward) events is therefore
$N_F\varepsilon+N_B
(1-\varepsilon)$ ($N_B\varepsilon+N_F(1-\varepsilon)$), where $N_{F(B)}$ is the true number of 
events in the forward region. As a result, the measured $A_{FB}^t$ equals 
$A_{FB}^{t~{\text{true}}}\times(2\varepsilon-1)$.  The suppression factor $\tilde\varepsilon$ defined 
in Eq.~(\ref{eq:ac_approx}) is shown on the right side of Fig.~\ref{fig:epsilon}.   Its value is 
\begin{equation}
\tilde\varepsilon=\frac{2\varepsilon^{LHC}-1}{2\varepsilon^{TEV}-1}\approx (54\pm10)\%.
\end{equation}
Combining all terms,  we expect  that  $A^t_C\simeq 0.17\times A_{FB}^{t}/88\% \times 
54\% \simeq 0.1\afbt$, where we recall that $\afbt$ is the value measured at the Tevatron.

Taking $\afbt \sim 20 \%$, we see that an extrapolation from the Tevatron provides 
a rough estimate for the LHC of 
$A_C  \simeq 0.02$, in reasonable agreement with the central value of the ATLAS 
measurement but in excess of the central value of the CMS measurement.   
Setting aside for the moment the still large uncertainties of the LHC data, the 
agreement of the ATLAS measurement with our extrapolation lends credence to 
the suggestion that new physics contributions are playing a role in the asymmetry 
measured at the Tevatron.  On the other hand, there is evident tension between 
the Tevatron asymmetry and the central value of the CMS measurement.  

Our model-based predictions of $A_C$, to be discussed presently, provide values 
of $A_C$  a little higher than the simple 
extrapolation.   The difference arises because the new physics contributions 
change the fraction of the $q\bar q$ initial state 
contribution to $t\bar t$ production at the Tevatron and the LHC.  The SM 
prediction for the $\bar{t} t$ cross section is 
$150\pm 19 \text{pb}$~\cite{mcfm}, 
and the ATLAS measurement is 
$177\pm3~({\rm stat.})^{+8}_{-7}~({\rm syst.})\pm 7~({\rm lumi.})~\text{pb}$ 
~\cite{ATLAS:2012tt5}.   
In the $W'$ and axigluon models, the contribution to $t\bar t$ production from new 
physics comes only through the $q\bar q$ initial state.  When the new physics 
contribution compensates for the excess of the measured $t\bar t $ cross section 
above the SM contribution,  the 
fraction from $q\bar q\to t\bar t$ to $t\bar t$ at $7~{\rm TeV}$ can increase to 
about $30\%$ compared with $17\%$ in SM. Therefore, we can expect $A_C 
\simeq 0.2 \afbt$, a factor of 2 enhancement with respect to our previous estimate.  

The analysis above provides an estimation of $A_C$ at the LHC from $\afbt$
at the Tevatron.  It should be used carefully as there are reasons that it may not 
be good enough. 
First, contributions from $t\bar t$ processes with extra partons in the final state 
are not included in the estimation.  They might be important for some new physics 
models especially for $A_C$~\cite{Alvarez:2012ca,Drobnak:2012rb,Adelman:2012py}.  
Second, there are models in which $\afbt$ at Tevatron is a residue of the balance
between contributions from $u\bar u$ and $d\bar d$ initial states~\cite{Drobnak:2012cz}.   
In this case, $A_C$ at the LHC could vary over a wide range 
since the fraction of the $u\bar u$ and $d\bar d$ initial states is different at the 
LHC,  and $\tilde\varepsilon$ for the $u$-quark and $d$-quark is different and dependent 
on the effective energy of the $t\bar t$ center of mass. 
Third, for new physics models in which the $\afbt$ results from a resonance effect,
there will be a suppression (enhancement) if the resonance is heavy (light
~\cite{Tavares:2011zg,Gresham:2012kv}).

\begin{figure}[!htb]
\includegraphics[scale=0.35]{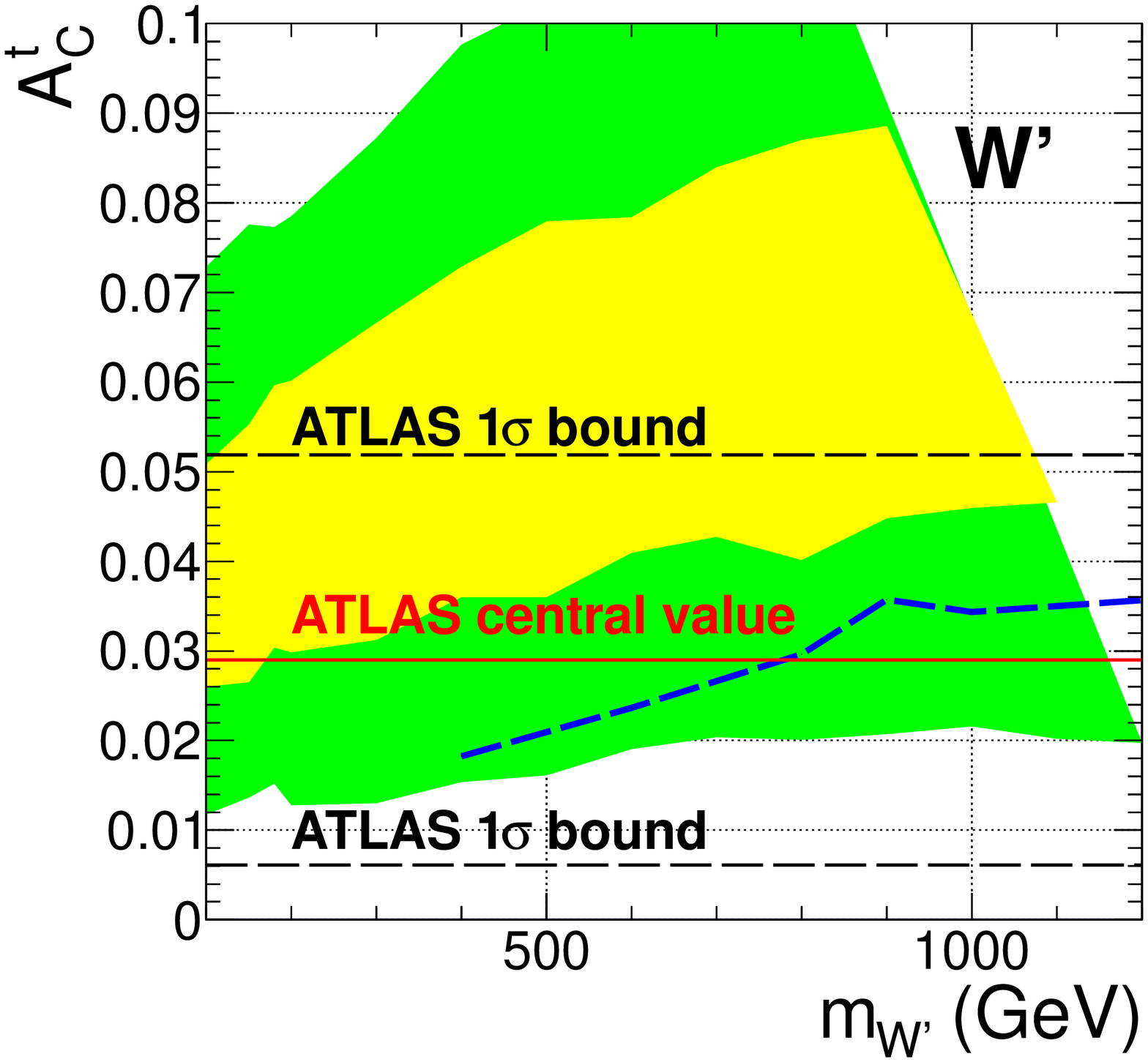}
\includegraphics[scale=0.35]{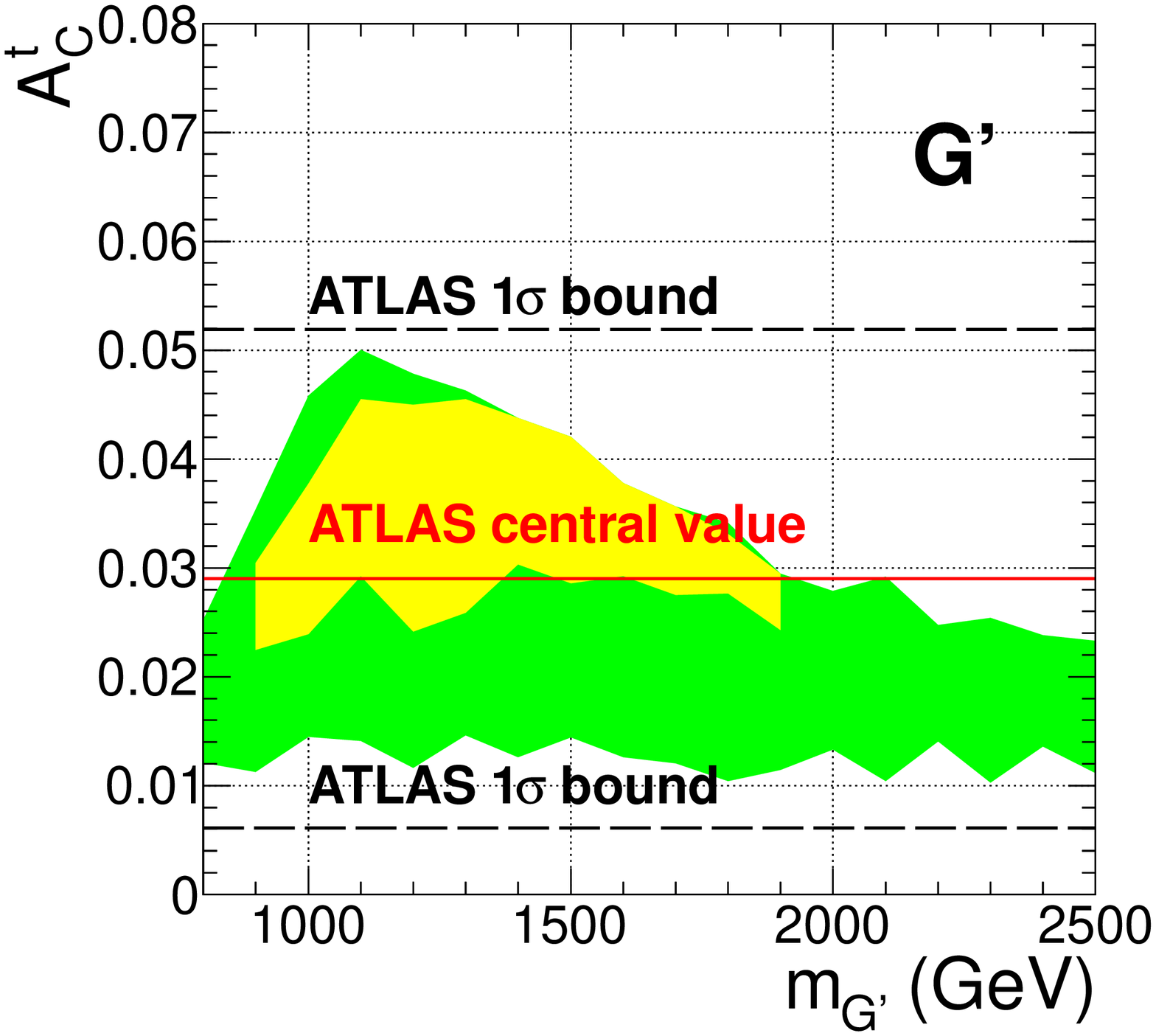}
\includegraphics[scale=0.35]{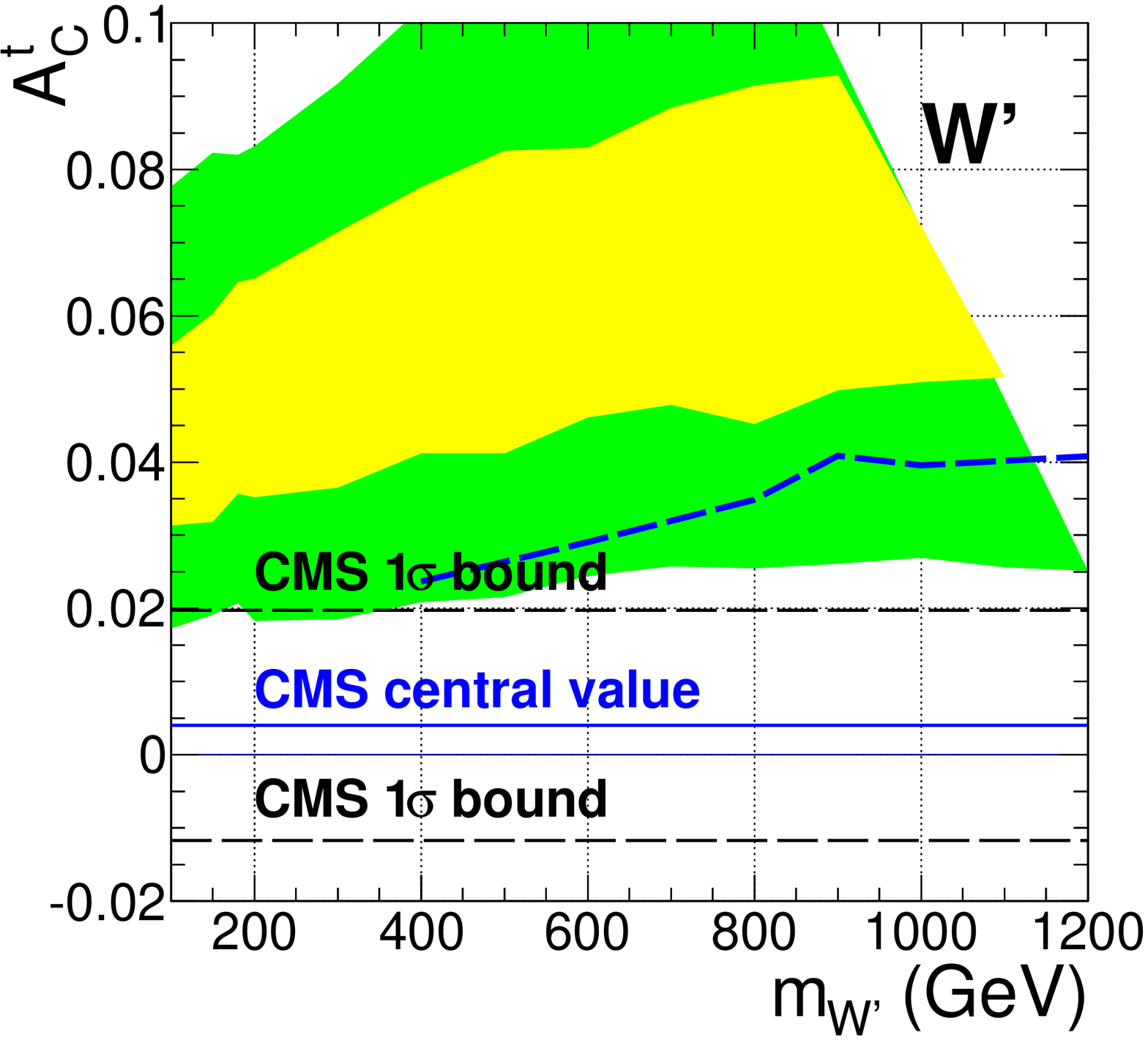}
\includegraphics[scale=0.35]{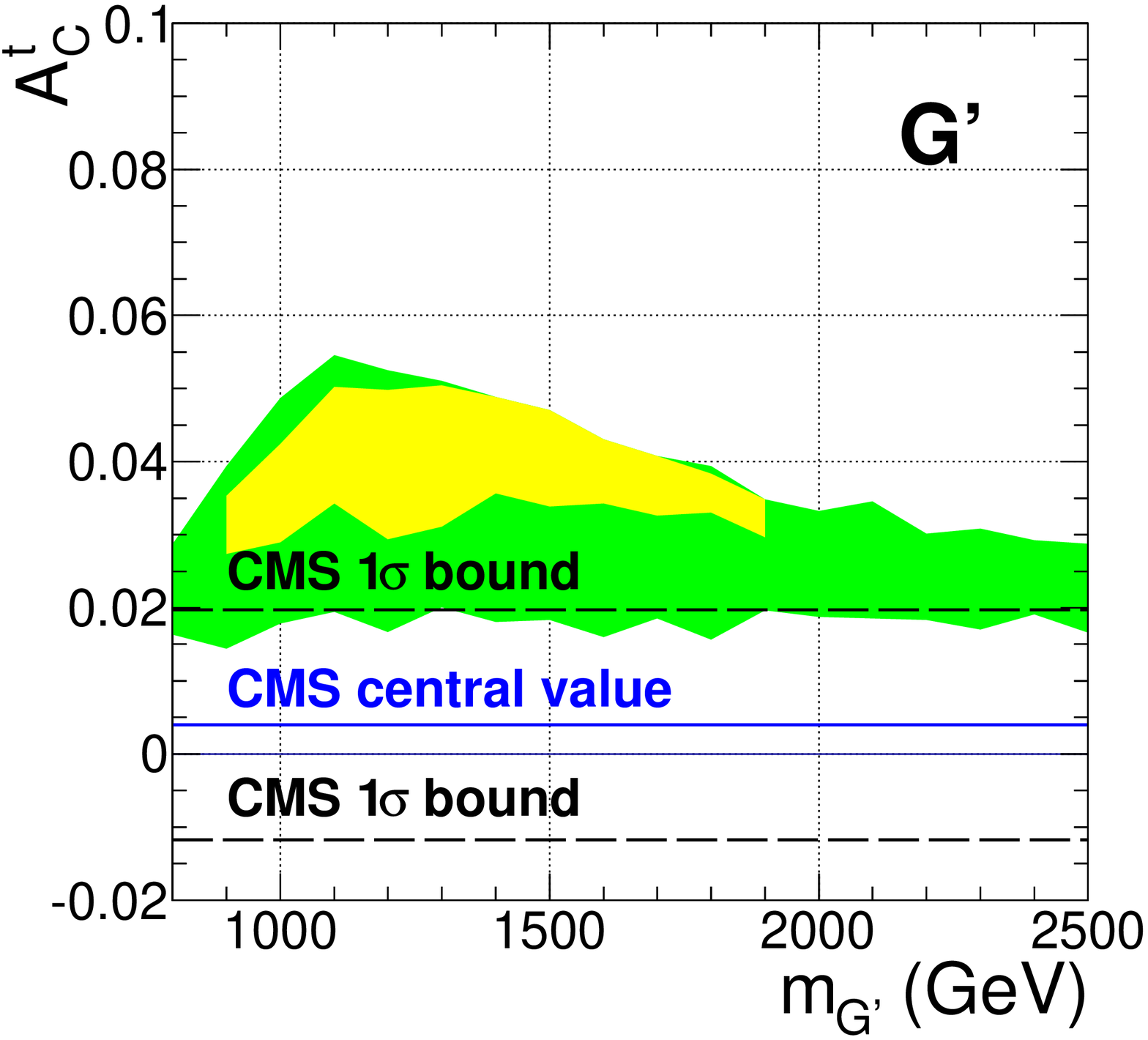}
\caption{The predicted top quark charge asymmetry, $A_C^t$, 
at the LHC at 7~TeV from the 
$W^\prime$ (left) and axigluon (right) models, compared with the ATLAS (upper) 
and CMS (lower) results.  The inner shaded (yellow) and outer shaded (green) regions are 
for the couplings that 
fit the Tevatron $t\bar t$ total cross section and  $\afbt$ within $1\sigma$ and 
$2\sigma $, respectively.  The central value measured by ATLAS (CMS) at the 
LHC is marked with a (red/blue) horizontal line, and the (black) horizontal dashed lines 
show the $1 \sigma$ uncertainty of the measurement.  The irregular (blue) dashed lines 
on the $W^\prime$ figures show the bounds obtained from the analysis of 
top-pair-plus-one-jet events.  The regions above these dashed lines are disfavored.}
\label{fig:lhcActop}
\end{figure}

Turning next to the explicit new physics models discussed in the previous section, we use 
the allowed parameters for the flavor-changing 
$W^\prime$ and axigluon models shown in FIG.~\ref{fig:tevfit} to calculate $A_C$  
at the LHC.   The results are shown in FIG.~\ref{fig:lhcActop},  along with a 
comparison to results of ATLAS and CMS.  We show different theory predictions 
for ATLAS and CMS.   The difference in the assumed value of the SM contribution explains the 
differences in the predictions of $A_C$ in FIG.~\ref{fig:lhcActop}.   To obtain the ATLAS 
predictions we use $A_C = 0.006$ for the SM prediction, as done by ATLAS.   For the CMS 
comparison, we use the SM value $A_C = 0.0115$ adopted by CMS.
  
The values $A_C$ for the $W'$ model at $7$ TeV are in the range $0.01~\sim 
0.1$.  The sharp drop 
for $m_{W'} = 850~{\rm GeV}\sim~1200~{\rm GeV}$ is related to our upper cut of 
the coupling parameter $g_{W'}$ at $7.5 $.  Most of 
the values of $A_C$ predicted in the $W'$ model are larger than the ATLAS 
central value; however, they are within the $1\sigma$ uncertainty band.   For the 
axigluon model, all of the predictions of $A_C$  agree with the ATLAS result within 
the $1\sigma$ level.  In the axigluon model $A_C$ does not simply increase with 
the axigluon coupling to SM particles.  For $m_{G'}=1500~{\rm GeV}$, $A_C$ 
reaches its maximum at about $4.2\%$, with coupling $f_{G'} =2.7$.  Therefore, 
we can see that the upper boundary of the yellow region (couplings that fit 
Tevatron data within $1\sigma$) overlaps the green region (couplings that fit 
Tevatron data within $2\sigma$) for some $m_{G'}$. 
The $G^\prime$ model predicts smaller values of $A_C$ than the $W^\prime$ model 
because there is a 
change of the sign of the {\it s-}channel propagator. When the invariant mass of 
the $t\bar{t}$ system is larger than the mass of the axigluon, the contribution to 
$A_C$ from the interference term is negative.    
In comparing with the CMS data, we see that owing to the large contribution from 
new physics, the predicted values of $A_C$ are outside of the $1\sigma$ band.  
Unless the central value increases in updated measurements, the CMS data 
disagree with new physics models based on $W'$ or axigluon contributions. 

For the $W'$ (and other $t-$channel new physics models), the associated production
process $d(\bar d)+g\to t(\bar t)+W'\to t\bar t+d(\bar d)$ may also give a significant 
contribution to $A_C$.   In~\cite{Alvarez:2012ca,Drobnak:2012rb}, such effects are 
investigated for a non-self-conjugate $Z'$ model.  The large gluon parton density 
accentuates the cross section for a relative light $Z'$ and $W'$, yielding a 
negative contribution to $A_C$~\cite{Alvarez:2012ca,Drobnak:2012rb} and releasing  
the tension between the small measurement at the LHC and the large
predictions from the new physics models.   The overlap between the predictions 
and the experimental bounds will be larger than shown in FIG. \ref{fig:lhcActop}.
However, a complete analysis must take into account interference between $t\bar t+j$ 
in the SM and the $W'$ model~\cite{Duffty:2012zz,Endo:2012mi} and a large enhancement 
from NLO QCD corrections ($K-$factor $\sim1.3-1.5$ \cite{Adelman:2012py}) .   
We defer it for future study.

\section{$\afbl$ and its correlation with $\afbt$}
\label{sec:landtasy}

In addition to the top quark forward-backword asymmetry, the charge lepton asymmetry 
$\afbl$ is also measured by the 
D0 collaboration at the Tevatron and by the ATLAS collaboration at the LHC.  It is defined as  
\be
A_{FB}^\ell = \frac{N_\ell^+ -N_\ell^-}{N_\ell^+ + N_\ell^-}.
\label{eq:def_afbl}
\ee
At the Tevatron, $N_\ell^+$ ($N_\ell^-$) is the number of events with $q_{\ell}y_{\ell} > 0$ 
$ (q_{\ell}y_{\ell} < 0)$, and $q_{\ell}$ and $y_{\ell}$ are the sign and rapidity respectively 
of the charged lepton from the semileptonic decay of a top or anti-top quark in the lepton 
plus jets events of  $t{\bar t}$ production. As stated in the Introduction, the 
D0 group reports $\afbl = (11.8 \pm 3.2)\%$, a deviation of about $2.2\sigma$ above the 
SM prediction $4.7\pm0.1\%$~\cite{:2012bfa}.
 At the LHC, the ATLAS collaboration measures $\afbl$ using data from the dilepton channel 
 in $t\bar t$ events;  $N_\ell^+$ ($N_\ell^-$) represents the number of events with 
 $|y_{\ell^+}| - |y_{\ell^-}| > (<) 0$.  Based on data corresponding to $4.7~{\rm fb}^{-1}$ of 
 integrated luminosity, ATLAS finds $(2.3\pm 1.2 ({\rm stat.}) \pm 0.8({\rm syst.}))\%$, in 
 excess of the SM prediction $0.4\%$ but 
 within $2\sigma$~\cite{ATLAS-CONF-2012-057}.  

The top quark is the only quark that decays quickly, before hadronization takes place, 
and its polarization determines the kinematic distribution of its final state particles.  
Therefore, it should be possible to understand $\afbl$ based on the kinematics of the 
charged lepton in the decay of a top quark with different polarization states.   Before 
presenting our numerical predictions for $\afbl$ in Sec~\ref{sec:scan2}, we show analytically 
how the relationship of $\afbl$ and $\afbt$ is controlled by the top quark polarization.
In this section, we start with the kinematics of a charged lepton in top quark decay 
and derive the correlation between $\afbt$ and $\afbl$. 
We introduce a variable $R_F$ that is useful for bridging the lepton asymmetry and 
the top quark asymmetry.   

\subsection{Lepton kinematics and top quark polarization}
\label{subsec:kin}

\begin{figure}[!htb]
\includegraphics[scale=0.3]{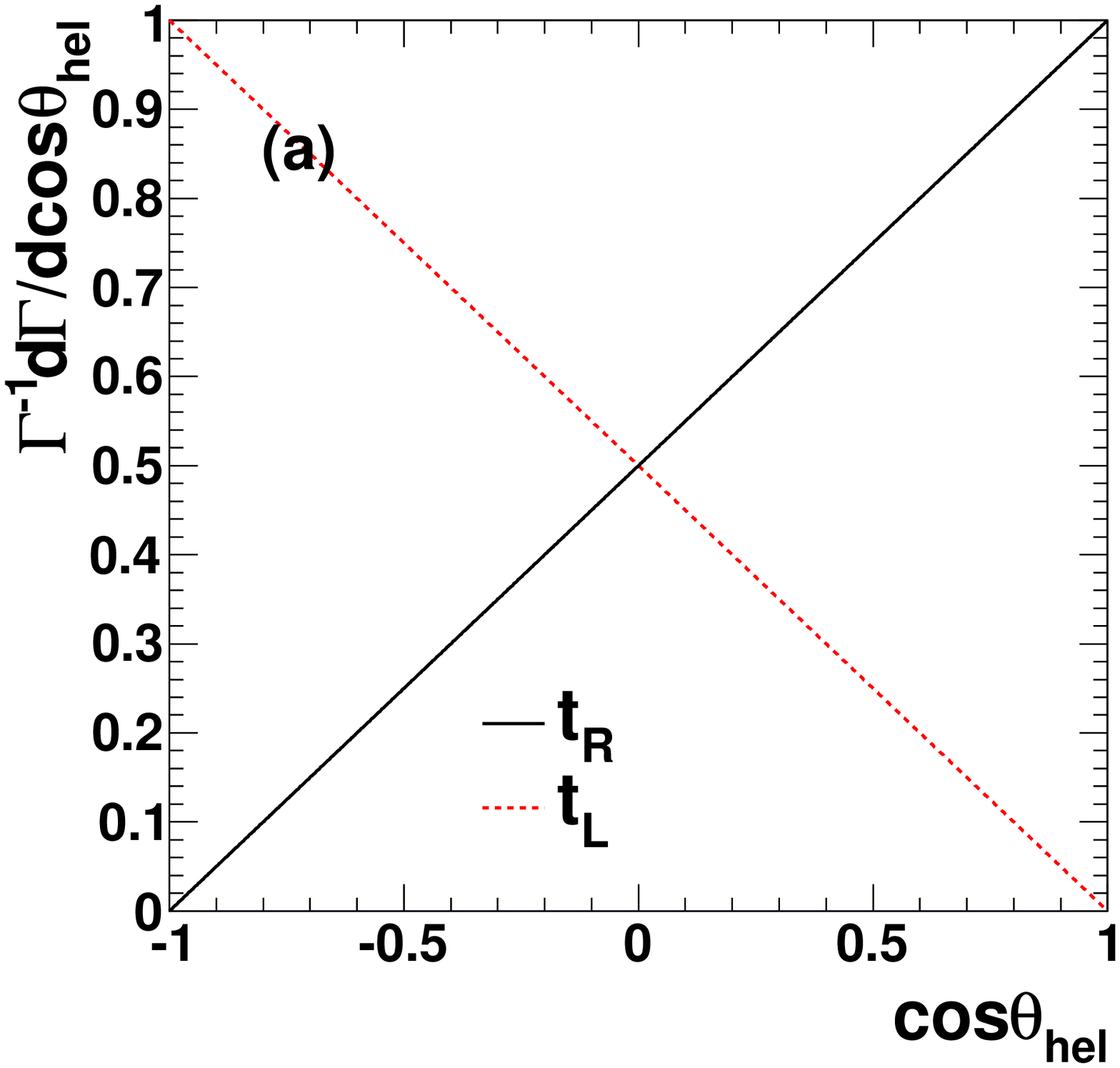}
\includegraphics[scale=0.3]{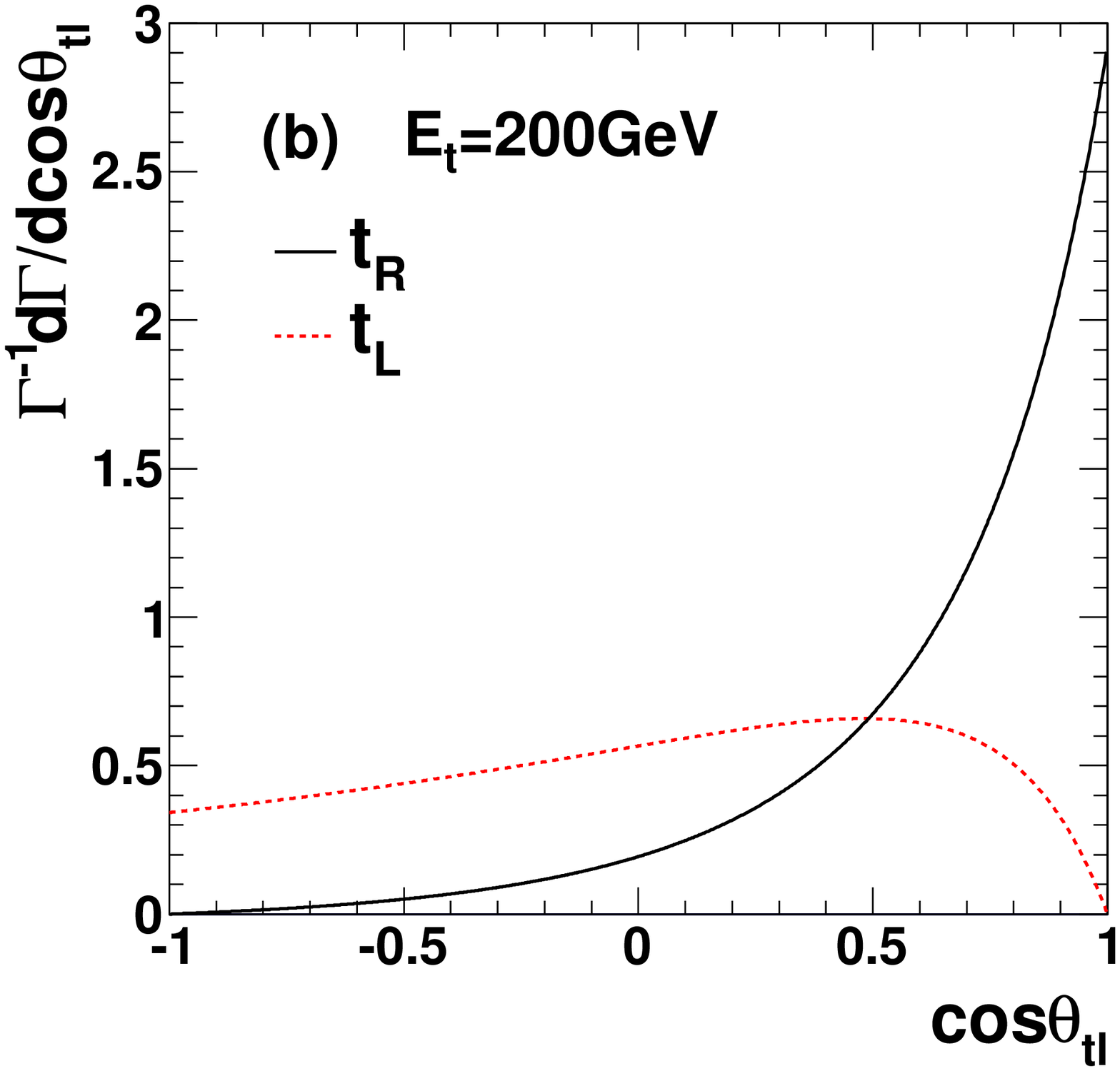}
\caption{(a) $\cos\theta_{\rm hel}$ distribution in the top quark rest frame for both $t_L$ 
and $t_R$. (b) $\cos\theta_{t\ell}$ distribution in the boosted frame for a top quark 
with $E_t=200~{\rm GeV}$.}
\label{fig:leprap}
\end{figure}

The charged lepton in top quark decay is a powerful analyzer of the polarization of 
the top quark~\cite{Mahlon:2010gw}. Owing to the $V-A$ structure of the charged 
current in the SM, the angular distribution of a charged lepton $\ell^+$ from top quark 
decay ($t\to W^+ (\to \ell^+ \nu) b$) in the top quark rest frame  is 
\be
\frac{1}{\Gamma}\frac{d\Gamma}{d\cos\theta_{\rm hel}}=\frac{1+\lambda_t\cos\theta_{\rm hel}}{2},
\label{eq:spin}
\ee
where $\lambda_t$ denotes the top quark helicity, and $\theta_{\rm hel}$  is the angle of $\ell^+$ 
with respect to the direction of motion of the top quark in the overall center-of-mass system of the 
$t \bar{t}$ production process.  
Throughout this paper, we use the helicity basis in our calculations.   We use $\lambda_t=+$  to denote 
a right-handed top quark ($t_R$),  and $\lambda_t=-$ for a left-handed top quark ($t_L$). 
The distributions are shown in FIG.~\ref{fig:leprap}(a).    
The charged lepton from a right-handed top quark decay prefers to move along
the top quark direction of motion, while a lepton from a left-handed top quark moves 
preferentially  against the top quark direction of motion.   In the rest frame of the top quark, 
75\% (25\%) of charged leptons from $t_R$ ($t_L$) decay follow the top quark direction 
of motion, i.e. $\cos\theta_{\rm hel} > 0$.  

Once the top quark is boosted along its spin direction, the angular distribution of the 
charged lepton relative to the 
direction of motion of the top quark deviates from $(1\pm \cos\theta)$, and it becomes 
sensitive to the energy of the top quark $E_t$ (or equivalently its velocity $\beta$).  
We derive    
\begin{equation}
\frac{d\Gamma}{\Gamma d\cos\theta_{t\ell}}=\frac{1-\beta\cos\theta_{t\ell}+\lambda_t
\left(\cos\theta_{t\ell}-\beta\right)}{2\gamma^2\left(1-\beta\cos\theta_{t\ell}\right)^3}, 
\label{eq:lep_follow_top}
\end{equation}
where $\beta=\sqrt{1-m_t^2/E_t^2}$, $\gamma=E_t/m_t$ and $\theta_{t\ell}$ is the 
angle between the charged lepton and the direction of motion of its parent top quark.
As an illustration, we plot in FIG.~\ref{fig:leprap}(b) the distribution of $\cos\theta_{t\ell}$
of the charged lepton for $E_t = 200$~GeV.   The leptons from both $t_L$ and 
$t_R$ move preferentially forward, more so for $t_R$ than $t_L$.   About $60\%$ of $\ell^+$ follow 
the top quark (i.e., $\cos\theta_{t\ell}>0$ ) for $t_L$, and almost $100\%$ for $t_R$. 

To obtain the forward-backward asymmetry in the laboratory frame, we must rotate the angular distribution 
in Eq.~\ref{eq:lep_follow_top} from the top direction of motion to the laboratory coordinate axes.  
We use a function $R_F^{\ell,~\lambda_t}(\beta, y_t)$ to represent the probability that a lepton 
with positive charge lands in the forward region when it originates from a top quark with velocity 
$\beta$, rapidity $y_t$, and polarization $\lambda_t$.    Formally, 
\be
R_F^{\ell,~\lambda_t}(\beta, y_t)=\frac{N_F^\ell}{N_F^\ell+N_B^\ell}.  
\label{eq:defR}
\ee
where $N_F^\ell$ ($N_B^\ell$) denotes the number of leptons $\ell$ in the 
forward (backward) region in the laboratory.   Moreover,
\be 
A_{FB}^{\ell,~\lambda_t}(\beta, y_t) = 2 R_F^{\ell,~\lambda_t}(\beta, y_t) - 1 .
\label{eq:defA}
\ee

It is noteworthy that an explicit analytic expression can be obtained for 
$R_F^{\ell,~\lambda_t}(\beta, y_t)$ in the laboratory frame.  
The derivation is somewhat lengthy, and it is presented in Appendix~\ref{sub:appRF}.  
We obtain 
\bea
R_F^{\ell,\lambda_t}(\beta, y_t)=\begin{cases}
\displaystyle \frac{1}{2}+\frac{1}{2\left(1+\gamma^{-2}\coth^2y_t\right)^{1/2}}+
\frac{\lambda_t\coth^2y_t}{4\beta\gamma^2\left(1+\gamma^{-2}\coth^2y_t\right)^{3/2}}
&  \displaystyle y_t \in \left[0,~~y_t^{\rm max}\right]\\
&\\
\displaystyle \frac{1}{2}-\frac{1}{2\left(1+\gamma^{-2}\coth^2y_t\right)^{1/2}}-
\frac{\lambda_t\coth^2y_t}{4\beta\gamma^2\left(1+\gamma^{-2}\coth^2y_t\right)^{3/2}},
& \displaystyle y_t \in \left[-y_t^{\rm max},~0\right]
\end{cases}
\eea
where 
\be
y_t^{\rm max}=\frac{1}{2}\ln\frac{1+\beta}{1-\beta}.
\ee

\begin{figure}[!htb]
\includegraphics[scale=0.3]{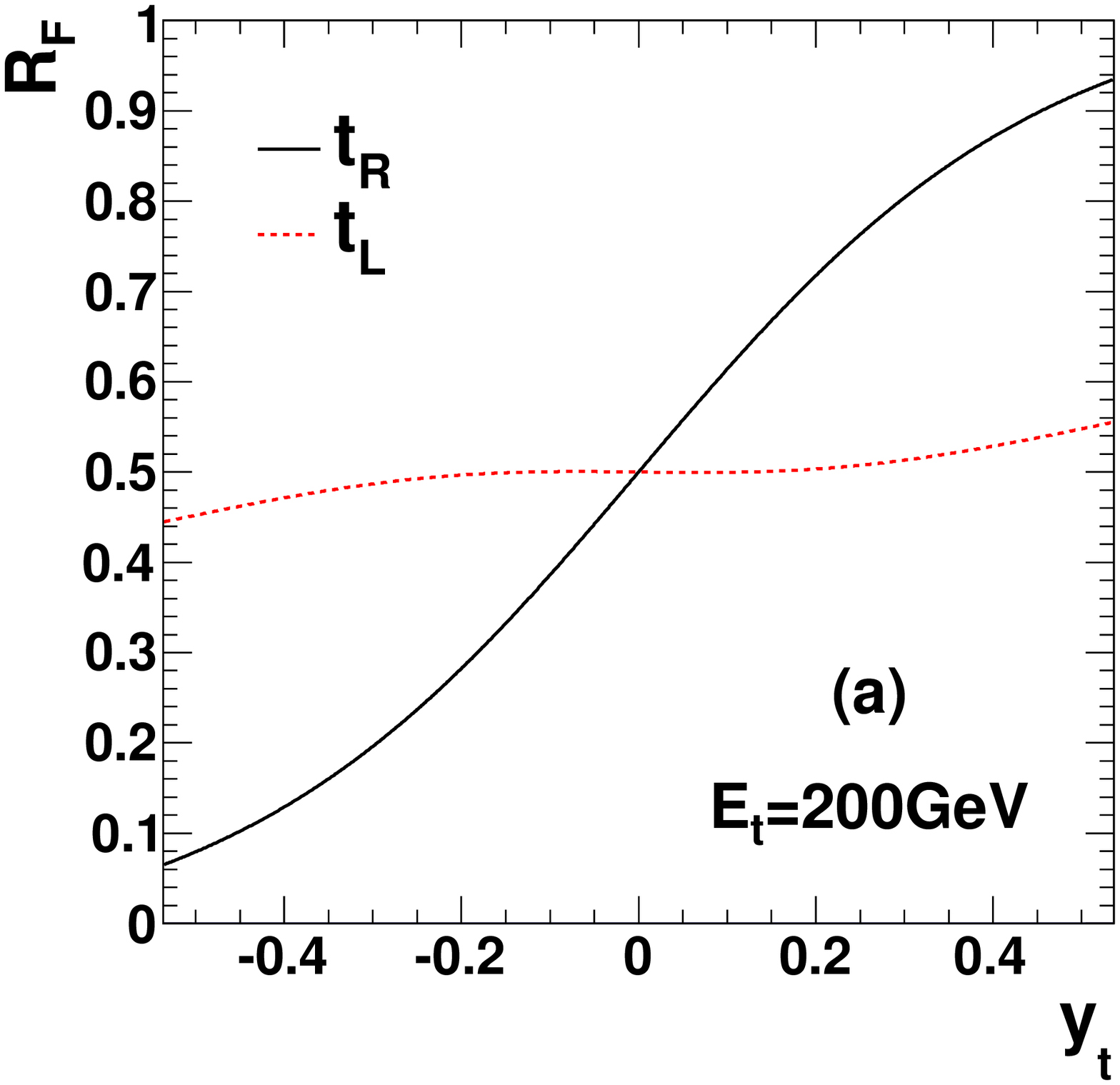}
\includegraphics[scale=0.3]{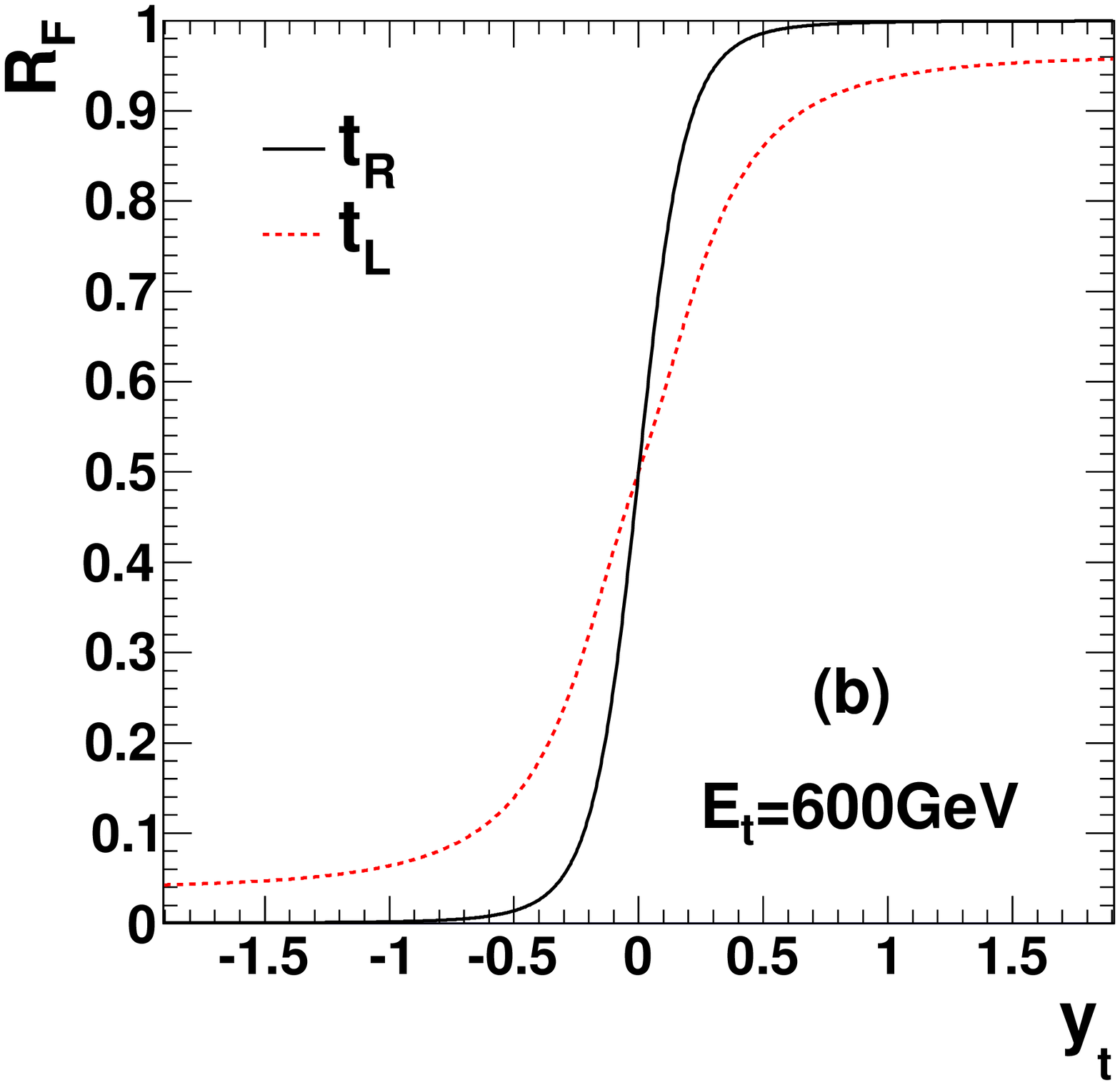}
\caption{
The ratio $R_F$ as a function of $y_t$ for a top quark with fixed energy: 
(a) $E_t = 200~{\rm GeV}$ and (b) $E_t=600~{\rm GeV}$.
\label{fig:lepratio2}}
\end{figure}

Figures illustrating the behavior of  $R_F^{\ell,\lambda_t}(\beta, y_t)$ as a 
function of $y_t$ for different choices of $E_t$, and as a function of $y_t$ 
for different choices of $p_t$ may be found in our Ref.~\cite{Berger:2012nw}, 
along with a discussion of interesting kinematic features of the curves. 
We limit ourselves here to showing FIG.~\ref{fig:lepratio2} and invite readers 
to consult our Ref.~\cite{Berger:2012nw}.  
The energy $E_t =200~{\rm GeV}$ represents top quarks produced just 
above the threshold region, where the cross section is greatest, while 
$E_t  = 600~{\rm GeV}$  pertains to highly boosted top quarks.   
For right-handed top quarks $t_R$ (black-solid lines in FIG.~\ref{fig:lepratio2}), 
$R_F$ increases rapidly with $y_t$ in the region $y_t  > 0$. 
On the contrary, in the case of $t_L$'s, the ratio $R_F$ does not vary 
as significantly with $y_t$. For $E_t = 200$ GeV, the boost causes charged leptons 
to distribute nearly uniformly, and as a result, $R_F$ is close to $0.5$ 
for the allowed range of $y_t$. When the energy of top quark is great 
enough, the large boost forces most the charged leptons from top quark 
decays to move along the top quark direction of motion, even for $t_L$.  

\subsection{From $A_{FB}^t$ to $A_{FB}^\ell$}
\label{subsec:scan1}

The functions $R_F^{\ell,~\lambda_t}(\beta, y_t)$ in Eq.~\ref{eq:defR} and 
$A_F^{\ell,~\lambda_t}(\beta, y_t)$ in Eq.~\ref{eq:defA} 
are functions of the top quark momentum.   To obtain the numbers of leptons 
in the forward and backward regions, we must convolve $R_F^{\ell,~\lambda_t}
(\beta, y_t)$ with the top quark momentum spectrum 
on an event-by-event basis, i.e. 
\begin{eqnarray}
\frac{N^\ell_F}{N^\ell_F+N^\ell_B}&=&\frac{1}{\sigma}\sum_{\lambda=+,-}
\int R_F^{\lambda}\left(\beta,y_t\right)
\frac{d^2\sigma|_{\lambda_t=\lambda}}{d\beta dy_t}d\beta \wedge dy_t,\\
\frac{N^\ell_B}{N^\ell_F+N^\ell_B}&=&\frac{1}{\sigma}\sum_{\lambda=+,-}
\int\left[1-R_F^{\lambda}\left(\beta,y_t\right)\right]\frac{d^2\sigma|_{\lambda_t=\lambda}}
{d\beta dy_t}d\beta \wedge dy_t,\\
A_{FB}^\ell&=&\frac{1}{\sigma}\sum_{\lambda=+,-}\int \left[2R_F^{\lambda}
\left(\beta,y_t\right)-1\right]\frac{d^2\sigma|_{\lambda_t=\lambda}}{d\beta dy_t}
d\beta \wedge dy_t
\label{eq:correlation}
\end{eqnarray}
where $\displaystyle{\frac{d^2\sigma|_{\lambda_t=\lambda}}{d\beta dy_t}}$ labels the 
differential $t\bar{t}$ production cross section for a top quark with specific kinematics
($\beta$, $y_t$, $\lambda_t$) and $\sigma$ stands for $t\bar{t}$ total
production cross section. 

The observed positive top-quark asymmetry $A_{FB}^t$ indicates that more top 
quarks are produced  in the forward region than in the backward region of rapidity.   
Both $t_R$ and $t_L$ can generate a positive lepton asymmetry $A_{FB}^\ell$ from 
a positive $\afbt$. However, a $t_L$ would need a large boost along the proton beam 
line (i.e. in the large forward rapidity region) to overcome the fact that most of the 
charged leptons from its decay move against it in its rest frame. A right-handed top 
quark $t_R$ can yield a positive $A_{FB}^\ell$ even for top quarks near the $t\bar{t}$ 
threshold region.  Therefore, the large positive top quark and lepton asymmetries 
$A_{FB}^t$ and $A_{FB}^\ell$ observed by the D0 collaboration indicate that
the top quark polarization and the kinematics of the top quarks, $y_t$ and $E_t$, may 
be playing a non-trivial role.   

In the SM, the vector coupling of gluons in the SM leads to equal 
production of left-handed and right-handed top quarks in the final state. After performing 
the convolutions in Eq.~\ref{eq:correlation}, we obtain  
\be 
\frac {\afbl} {\afbt} \simeq \frac {(0 + 0.8)} {2} = 40\%
\ee
in the SM at Tevatron.   The first term in the numerator is the contribution from left-handed 
top quarks, and the second term is from the right-handed top quarks.   This estimate agrees 
well with explicit NLO calculations~\cite{Abazov:2011rq}.  

This SM expectation may be contrasted with the value 
\be 
\frac {\afbl} {\afbt} =  (78 \pm 33) \%
\ee
obtained from the D0 measurements of $\afbt$ and 
$A_{FB}^\ell=(15.2\pm 4.0)\%$ measured in the {$\ell$+jets} final states~\cite{Abazov:2011rq}. 
On the other hand, using the value $A^{\ell}_{\rm FB}= (11.8 \pm 3.2)\%$ obtained from a 
combination of measurements in the dilepton final states from $t \bar{t}$ production and the 
$\ell$+jets final states, we find 
\be 
\frac {\afbl} {\afbt} =  (60 \pm 26) \%
\ee
The uncertainties are large, but the central values of these ratios exceed the SM estimate and 
indicate that the physics responsible for the forward-backward asymmetry produces more 
right-handed than left-handed top quarks.   It would be valuable to 
confirm the measurement of $A_{FB}^\ell$ with the full data sample in D0 and to make a 
similar measurement with CDF data.

The top quark asymmetry $A_{FB}^t$ can be expressed as a sum of contributions from the 
SM and NP as: 
\bea
A_{FB}^t &=& \frac{(N_F^{\rm  NP}+N_F^{\rm  SM})-(N_B^{\rm  NP}+N_B^{\rm  SM})}
{(N_F^{\rm  NP}+N_F^{\rm  SM})+(N_B^{\rm  NP}+N_B^{\rm  SM})} \nonumber \\
               &=& A_{FB}^{t,~{\rm  NP}}\times R^{\rm NP} + A_{FB}^{t,~{\rm SM}}\times (1-R^{\rm NP}),
\eea
where 
\be
A_{FB}^{t,~{\rm SM}}=\frac{N_F^{\rm SM}-N_B^{\rm SM}}
{N_F^{\rm SM}+N_B^{\rm SM}},~~
A_{FB}^{t,~{\rm NP}}=\frac{N_F^{\rm NP}-N_B^{\rm NP}}
{N_F^{\rm NP}+N_B^{\rm NP}},~~
R^{\rm NP}= \frac{N_{\rm tot}^{\rm NP}}{N_{\rm tot}^{\rm SM}+N_{\rm tot}^{\rm NP}},
\ee
with $N_{F(B)}^{\rm SM}$ and $N_{F(B)}^{\rm NP}$ being the numbers of events in which the 
top quark moves with $y_t>0 (y_t<0)$ in the SM and induced by NP, respectively, and 
$N_{\rm tot}^{\rm SM (NP)}$ is the total number of events predicted in the SM (induced by NP).

A simplified analysis the correlation between $A_{FB}^t$ and $A_{FB}^\ell$ in presented 
in our Ref.~\cite{Berger:2012nw} in which we assume that $A_{FB}^t$ is 
generated entirely by new physics. 
In the explicit numerical predictions presented in the next section all SM contributions 
including the NLO QCD effects are retained. 

\section{$\afbl$ and new physics models: axigluon and $W^\prime$}
\label{sec:scan2}

\begin{figure}[!htb]
\includegraphics[scale=0.35]{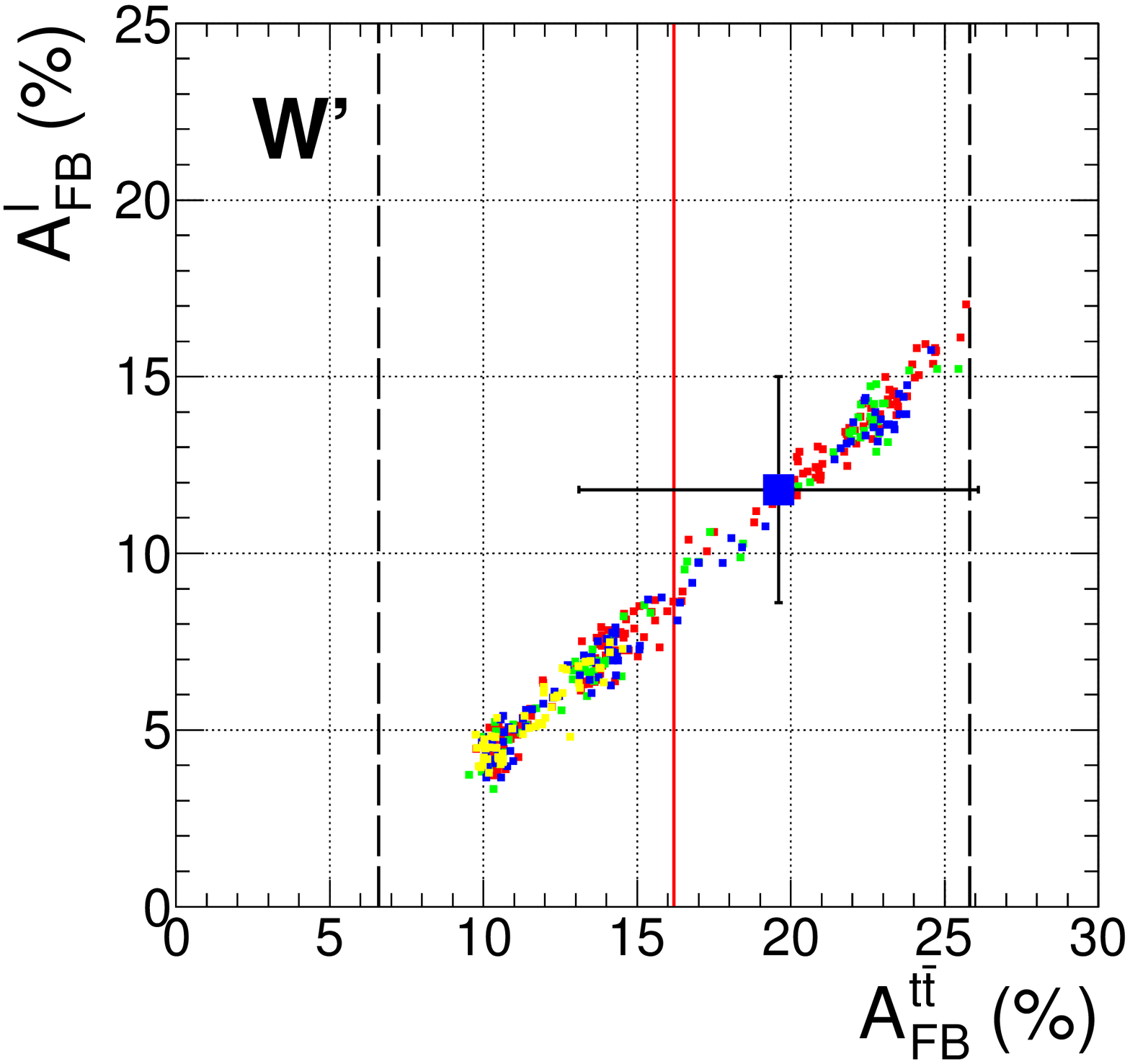}
\includegraphics[scale=0.35]{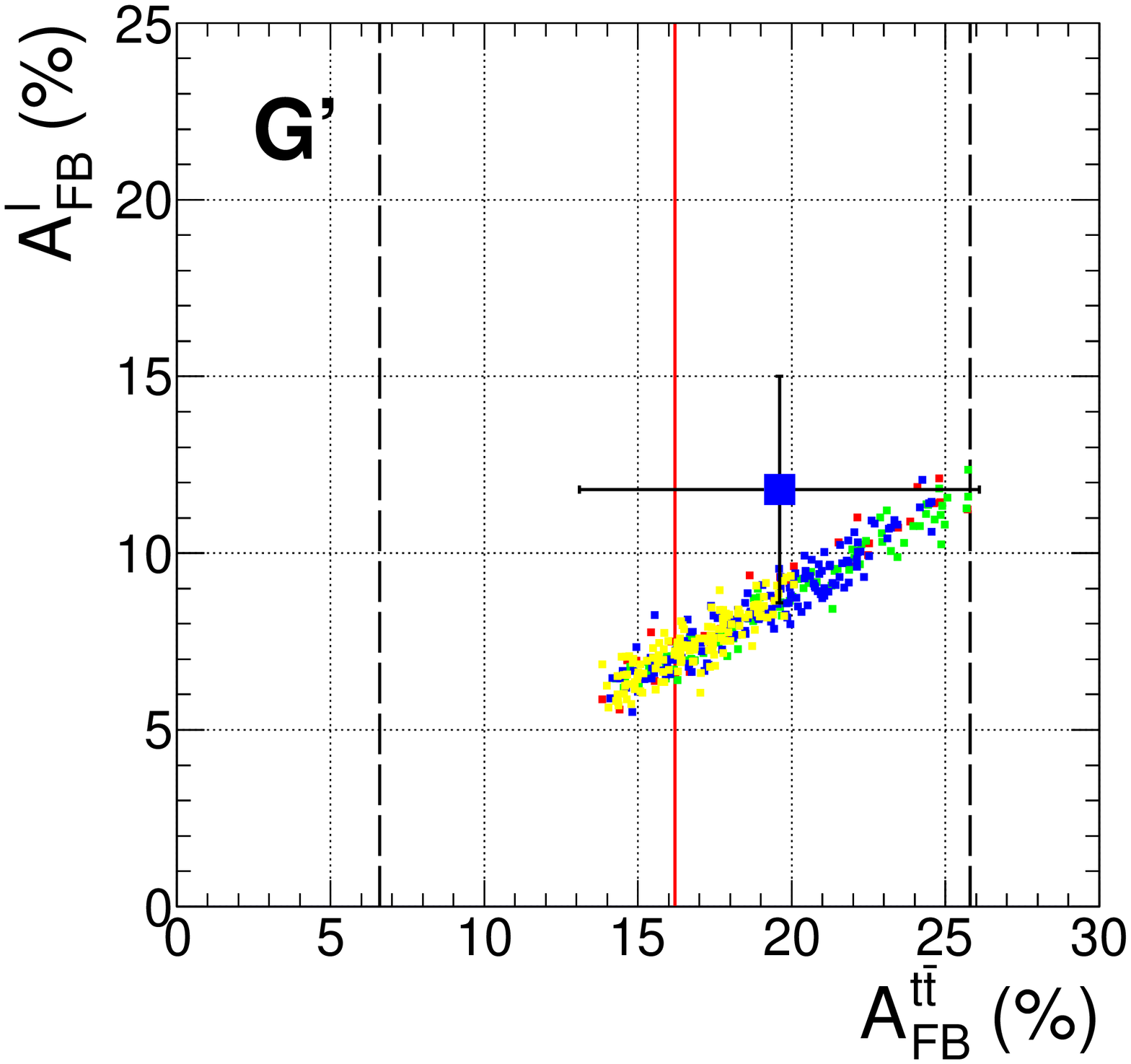}
\includegraphics[scale=0.35]{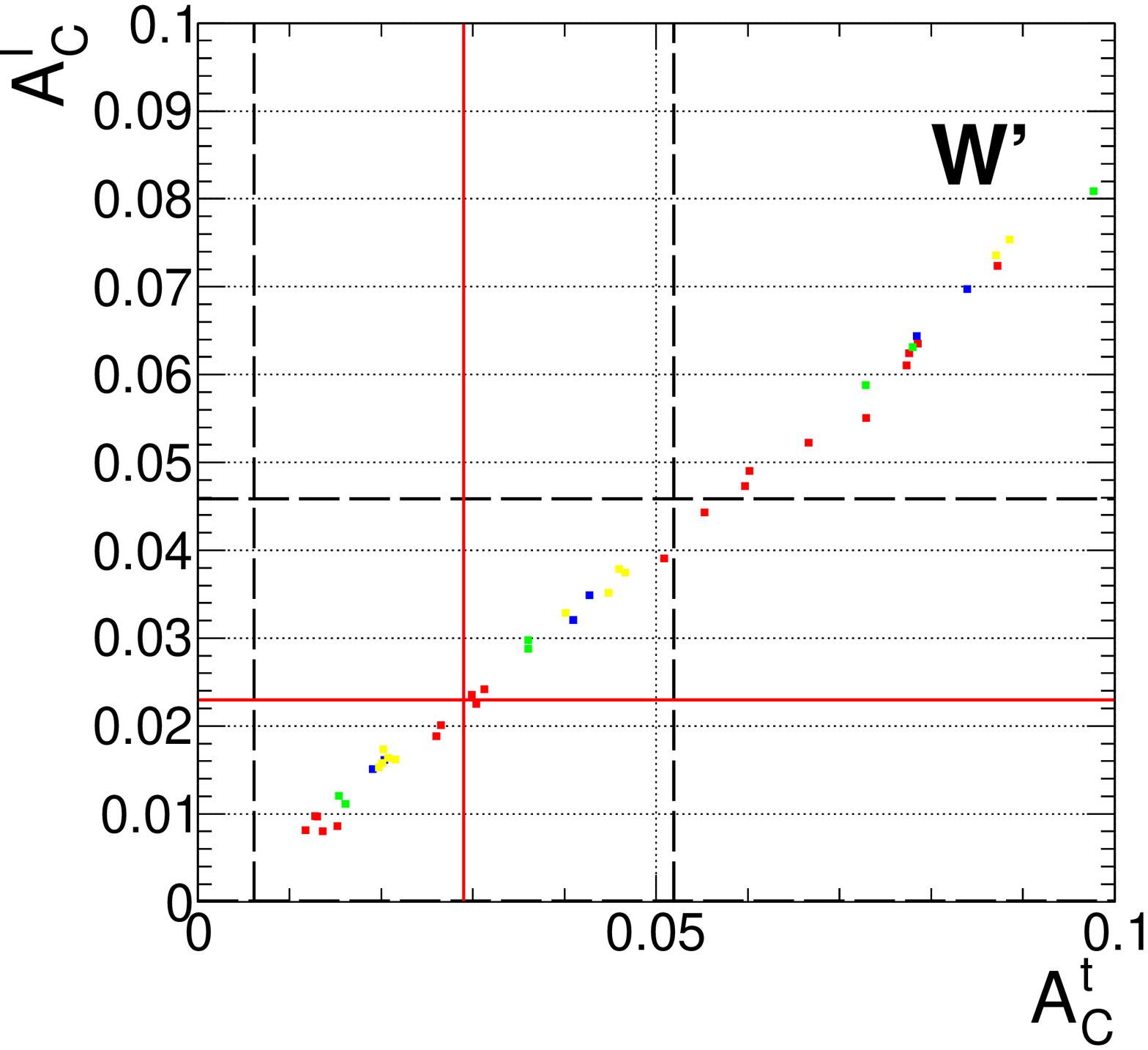}
\includegraphics[scale=0.35]{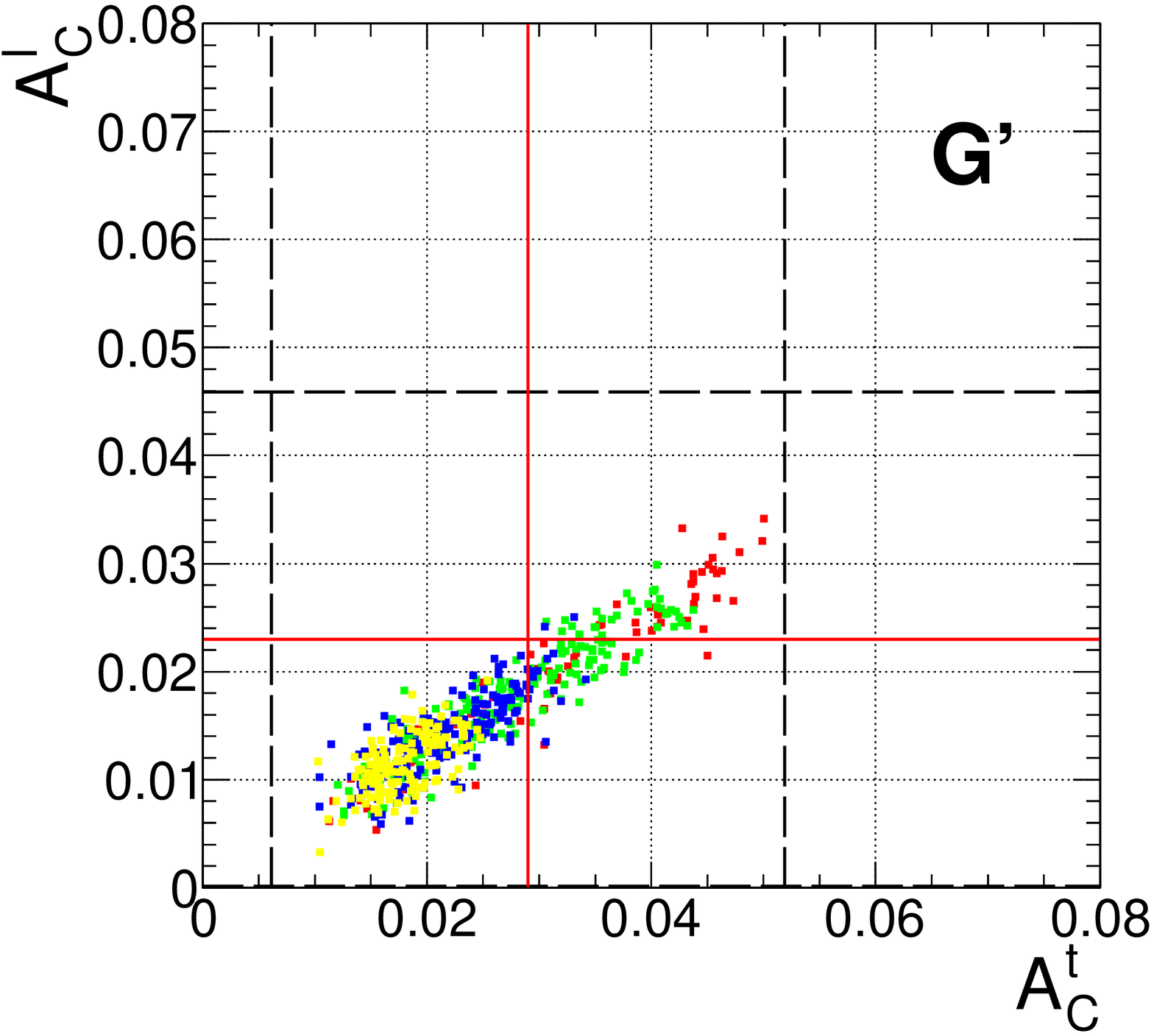}
\caption{Upper two panels: correlation between $A_{FB}^{\ell}$ and $A_{FB}^{t}$ 
for the $W^\prime$ (left) and axigluon (right) models at the Tevatron.  Different colors 
of dots indicate different mass ranges.   In the $W'$ model, red: $100~{\rm GeV}\sim 
300~{\rm GeV}$, green: $400~{\rm GeV}\sim 500~{\rm GeV}$, blue: $600~{\rm GeV}
\sim 800~{\rm GeV}$, yellow:$>800~{\rm GeV}$.  In the axigluon model,  red: 
$800~{\rm GeV}\sim 1300~{\rm GeV}$, green: $1400~{\rm GeV}\sim 1700~{\rm GeV}$, 
blue: $1800~{\rm GeV}\sim 2100~{\rm GeV}$, yellow:$>2200~{\rm GeV}$. The (blue) 
square point is the D0 data $A^{\ell}_{\rm FB}= {11.8 \pm 3.2}\%$ with its  $1\sigma$ 
uncertainty.  The (red) vertical line shows 
$\afbt$ measured by CDF in the $8.7~{\rm fb}^{-1}$ data set, and the two vertical (black) 
lines present the $1\sigma$ uncertainty band.  Lower two panels: the correlation between 
$A_C^t$ and $A_C^\ell$ at the LHC for the $W^\prime$ (left) and axigluon models 
(right).  The vertical (horizontal) red line and the two black dashed lines show the central 
value of $A_C^t$ ($A_C^\ell$) and the $1\sigma$ uncertainty bands measured by ATLAS 
at the LHC.}
\label{fig:correlation}
\end{figure}

The correlation between the charged lepton asymmetry and the top quark 
asymmetry is significantly different for different polarization states of the top 
quark, and it may therefore shed light on the nature of the physics that 
causes the forward-backward asymmetries at the Tevatron.   In this section, 
as in our previous study~\cite{Berger:2012nw}, we choose the $W^\prime$ 
and axigluon models as two reference models to examine the correlation at 
the Tevatron and the LHC.   The results we show here for the Tevatron are 
slightly different from our previous results because we now use parameters 
obtained in Sec~\ref{sec:update} from our fit to the CDF $8.7 {\rm fb}^{-1}$ 
data set.  In addition, we present predictions for the LHC.  

The axigluon and  $W^\prime$ models admit good fits to $\afbt$ at the Tevatron, 
but they provide distinct predictions for the polarization and kinematics of the final 
state top quark.  The $W'$ model produces dominantly $t_R$ while the axigluon 
model generates an equal number of $t_R$ and $t_L$ with more energetic top 
quarks since the quarks come from the decay of a heavy axigluon. In FIG.~\ref{fig:correlation}, 
we show the results of our calculation of the charged lepton asymmetry using the parameters 
determined in our $1\sigma$ fits to the $t\bar t$ total cross section and the most recent 
CDF data on $\afbt$ ($\afbt(m_{t\bar t}<450{\text{GeV}}),~\afbt(m_{t\bar t}
>450{\text{GeV}})$ and $\afbt$).  The upper two plots show the charged lepton asymmetry as a 
function of the top quark asymmetry at the Tevatron.   The lower two plots display the 
charged lepton asymmetry for the LHC together with the top quark charge asymmetry 
$A_C^t$.    For the Tevatron, the values of $\afbt$ are determined in the $t\bar t$ rest 
frame whereas, for comparison with the D0 point shown, $\afbl$ is in the laboratory frame.
For the LHC predictions, both at $A_C^t $ and at $A_C^l$ are in the same frame.  

There are vertical red lines in FIG.~\ref{fig:correlation} at $\afbt \sim 16\%$ in the 
Tevatron plots and at $A_C^t \sim 0.03$ in the LHC plots to show the central values 
of the asymmetries measured by CDF and ATLAS, respectively, and two black dashed 
lines in the upper and lower plots to show the extent of the quoted experimental 
$1\sigma$ uncertainty bands.   The horizontal red line in the LHC plots shows the central 
value of $A_C^\ell$ measured by ATLAS at the LHC, and the horizontal black dashed lines 
show the $1\sigma$ uncertainty values.    Since the CDF collaboration does not present 
the charged lepton asymmetry $A_{FB}^\ell$, we show only the D0 data 
$A^{\ell}_{\rm FB}= (11.8 \pm 3.2)\%$ as a blue square point.   

The calculated charged lepton asymmetries stretch out over a range of values depending 
on the values of the axigluon or $W'$ masses used in the fits to the Tevatron data.  At the 
Tevatron, the charged lepton asymmetry spreads from $3\%$ to $17\%$ in $W'$ the model, 
and over a narrower range, from $6\%$ to $13\%$ in the axigluon model.   
The D0 data point is in agreement with both models within uncertainties.
At the LHC, 
there are parameters in both models (obtained from the Tevatron fits) that can reproduce 
the values of $A_C^t$ and $A_C^\ell$ measured at the LHC by ATLAS, shown by the fact 
that the intersection of the vertical and horizontal red passes through the scattering of dots. 
On the other hand, there is a wide range of dots in the $W'$ model that are above the central 
values of $A_C^t$ and $A_C^\ell$, and out of the $1\sigma$ uncertainty band.  In the 
axigluon model, all the values of $A_C^t$ and $A_C^\ell$ are consistent with ATLAS 
measurements within the $1\sigma$ bands.   It is evident that LHC and Tevatron data 
together could reduce the allowed parameter spaces of the two models.  

The best fits to the lines of points in FIG.~\ref{fig:correlation} at the Tevatron are
\be
\afbl = 0.77\times\afbt-3.6\%~(W');~~\afbl = 0.50\times\afbt-1.0\%~({\rm axigluon}).
\ee  
For LHC, the best fits are
\be
A_C^\ell = 0.85\times A_C^t-0.002~(W');~~A_C^\ell = 0.61\times A_C^t+0.0008~({\rm axigluon}).
\ee  
\begin{figure}[!htb]
\includegraphics[scale=0.4]{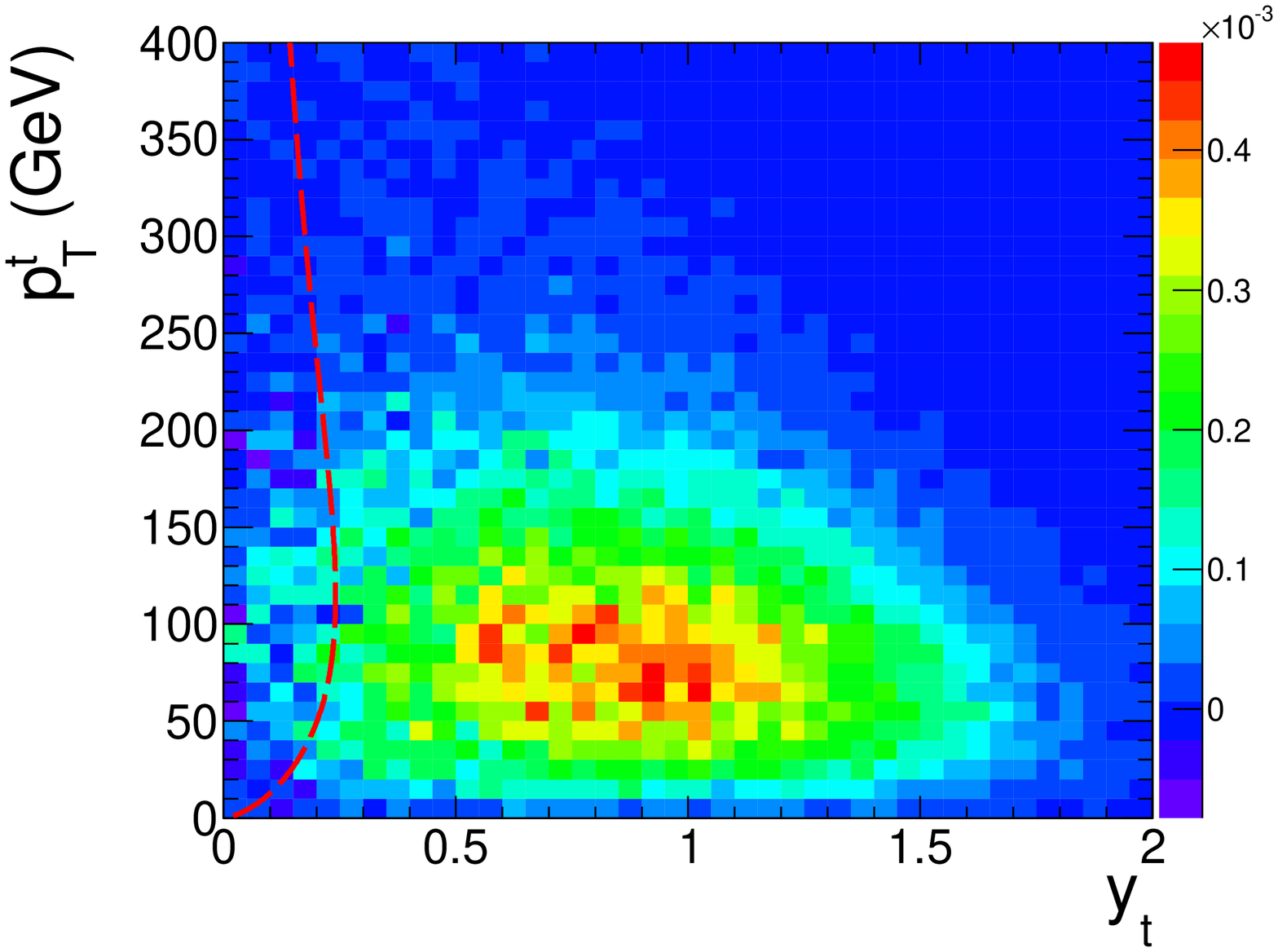}
\includegraphics[scale=0.4]{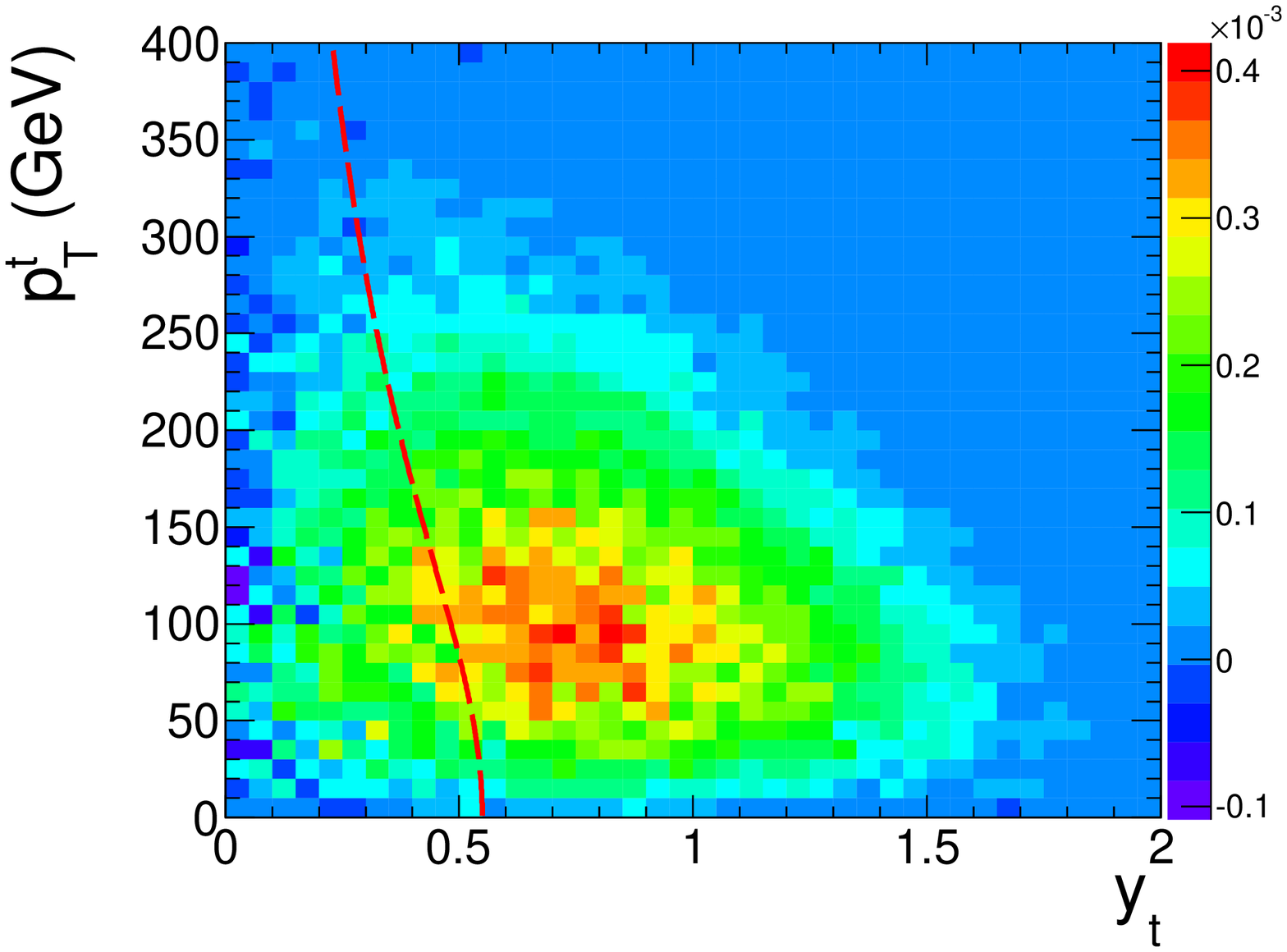}
\caption{Two-dimensional differential distribution, $(N^t(p_t,y_t)-N^t(p_t,-y_t))_{y_t >0}
/N_{\rm tot}$ for (a) the $W^\prime$ model with $m_{W^\prime}= 200$ GeV,  and (b) 
the axigluon model with $m_{G^\prime}= 1.8$ TeV.  The dashed line  indicates 
$2R_F -1 = 0.5$ for each model, with $2R_F -1 > 0.5$ $ (<0.5)$ on the right- (left-) side 
of the curve. The coupling strengths in the new physics models are chosen so that 
$\afbt$ and $\afbl$ are accommodated.}
\label{fig:2dpty}
\end{figure}

In order to gain greater insight into these correlations, we examine two-dimensional 
differential distributions of $A_{\rm FB}^t=(N_F^t-N_B^t)/N_{\rm tot}$ as a function of 
top quark rapidity and transverse momentum. In FIG.~\ref{fig:2dpty} we show these 
density plots for a $200$ GeV $W^\prime$ (left) and a $1.8$ TeV $G^\prime$ (right) 
at the Tevatron.  Different colors show different densities of $(N^t(p_t,y_t)-N^t(p_t,-y_t))_{y_t >0}
/N_{\rm tot}$.   The top quark forward-backward asymmetry $\afbt$ is obtained after 
integrating over the rapidity $y_t$ and transverse momentum ($p_T^t$) of the top quark. 
As we can see, most of the events which contribute to the top quark asymmetry concentrate 
in the region of $p_T$ about $50~{\rm GeV} \sim 150~{\rm GeV}$ (the axigluon model 
has more events with high $p_T$), and $y_t$ about $0.4\sim 1.2$. 

In FIG.~\ref{fig:2dpty}, we also show the curve of $2R_F -1 = 0.5$ as a red dashed curve.  
Events to the right (left) of the curve denote values $2R_F -1 >(<) 0.5$.   Note that  
$2R_F -1$ is the weight when we convolute with the differential $\afbt$ to obtain the 
charged lepton asymmetry, c.f. Eq.~(\ref{eq:correlation}). Therefore, a larger charged 
lepton asymmetry is expected when there are more events to the right of the red dashed 
curve.   In the $W^\prime$ model, events that contribute to $\afbt$ are more concentrated 
in the region $2R_F -1>0.5$ than for the axigluon model, consistent with the fact that 
$\afbl/\afbt$ in the $W'$ model is larger than in the axigluon model.   

The size of the top quark asymmetry, in excess of SM expectations,  is one indication 
that new physics may be playing a role. The charged lepton asymmetry provides a second 
and independent indication of the presence of new physics since 
it points toward the possibility that more right- 
than left-handed top quarks are being produced.   It is important to confirm the charged 
lepton asymmetry.   This goal could be realized with an analysis of the full data set in D0.  
We also encourage the CDF collaboration to measure the charge lepton asymmetry.

\section{Summary and Discussion}
\label{sec:con}

A forward-backward asymmetry in rapidity $\afbt$ of top quarks is observed at the Fermilab 
Tevatron.  Its value exceeds SM expectations by $\sim 3 \sigma$, perhaps indicative of the 
presence of new physics contributions in the top quark sector. In this paper we expand 
considerably on our previous studies of implications of the asymmetry and include new predictions 
for the related top quark charge asymmetry at LHC.   Starting from the CDF value of $\afbt$ 
obtained in the analysis of their $8.7~\rm{fb}^{-1}$ data set,  we derive the allowed regions 
parameter space of three illustrative new physics models, based, in turn, on the exchange of a flavor-changing 
heavy $Z'$ or $W'$ in the $t$-channel, or the contribution of an axigluon $G'$ in the $s$-channel. 
The asymmetry data alone now show that the minimal $Z'$ model is disfavored, a conclusion 
reinforced by the negative search by CMS for pairs of like-sign top quarks at the LHC.  For the 
$W'$ and $G'$ models, we show that the parameter space allowed by the asymmetry data is 
constrained further by LHC searches for $t\bar {t} $~plus one jet events and for enhancements in 
the dijet mass distribution, respectively. More sophisticated models can certainly 
be devised as extensions of the simple $Z'$, $W'$, and axigluon $G'$ models considered here.  
Our conclusions are limited to the models defined in Sec.~\ref{sec:update}.

Our analysis of the Tevatron data is then used to obtain predictions for $A_C$ at the LHC. 
First, the association of the asymmetry $\afbt$ with the quark-antiquark initial state allows 
us, by an extrapolation in energy, to obtain an estimation of 
$A_C \simeq 0.1 \afbt$, in agreement with the 
central value of the ATLAS measurement but in excess of the central value of the CMS data.  
Explicit calculations of $A_C$ based on the allowed parameter space of the $W'$ and $G'$ 
models are shown in  Fig.~\ref{fig:lhcActop} and compared with the LHC measurments by 
ATLAS and CMS.  These calculations confirm that it is difficult to reconcile the 
CMS measurements of $A_C$ with the parameters determined from fits to $\afbt$ at the 
Tevatron.  On the other hand, the ATLAS data are readily accommodated. The available LHC 
data on $A_C$ are based on a sample with only $4.7~\rm{fb}^{-1}$ of integrated lumiosity. 
A reduction of the experimental uncertainties could justify stronger conclusions regarding 
the compatibility of the Tevatron and LHC measurements, and a combined analysis of full 
statistics data from both colliders would offer significant advantages.  

As discussed in Sec.~\ref{sec:update}, we fit Tevatron data on the inclusive total cross section 
for $t\bar t$ production and $\afbt$ in order to determine the parameters of the new physics 
models under consideration, explaining the reasons we do not include  
data on the differential cross section in the invariant mass $m_{t \bar{t}}$ (see, in particular, the 
paragraph immediately 
following  Eq.~(\ref{eq:sm_afbt}) and the next-to-last paragraph of the same section).  More recent 
measurements of  the $m_{t\bar t}$ distribution at the LHC by the ATLAS~\cite{ATLAS:2013aa} and
CMS~\cite{CMS:2013aa,CMS:2013ab} collaborations invite consideration of a different approach from 
ours, in which data from the Tevatron and the LHC are used in a joint fit to determine model 
parameters.  The inclusion of differential data could provide further 
constraints on the allowed parameters of models of new physics.  No excess beyond the prediction of 
the SM is observed in the region of large $m_{t\bar t}$ in the LHC data, suggesting stringent limitations on 
models that predict an increase in the $t\bar t$ rate at high $m_{t\bar t}$.  This constraint is 
investigated in Refs.~\cite{AguilarSaavedra:2011ug,Aguilar-Saavedra:2013rza} where the cross 
section $\sigma\left(m_{t\bar t}>1{\text{TeV}}\right)$ is required to remain within 50\% of its SM value.  
The $Z'$ model is shown to be further excluded by this requirement, while the $W'$ model is constrained 
with a tiny positive contribution to $A_{FB}^t$.  For an $s$-channel axigluon model, our results in 
Sec.~\ref{sec:update} show that dijet searches at the LHC exclude a narrow width axigluon whose
mass is in the range [800 GeV, 2500 GeV].  Moreover, as we mention, there are subtleties in the 
use of the $m_{t\bar t}$ distribution in attempts to constrain a $G'$ with broad width.

Once statistical precision improves sufficiently at large values of  $m_{t \bar{t}}$, there is no doubt that 
fits to the differential distribution in $m_{t \bar{t}}$ should be done.   However, we caution again that 
a thorough analysis would require computation of the new physics contributions at NLO, include the 
effects of parton showering, and take into account experimental acceptance cuts whose effects are 
particularly significant at large values of $m_{t \bar{t}}$ (c.f., Ref~\cite{Gresham:2011pa}).  The analysis 
in terms of new physics models is also complicated by the fact that data on the $m_{t \bar{t}}$ distribution 
are unfolded in terms of the SM shape and cut efficiencies.  When considering  models 
more sophisticated than those we use here for illustrative purposes, one should bear in mind that the 
ultraviolet (UV) completion of the effective model can include the introduction of new particles that affect 
the reliable prediction of the large mass tail of the $m_{t\bar t}$ distribution (see, for example 
Ref.~\cite{Ko:2011vd,Ko:2011di,Ko:2012gj,Ko:2012ud,Ko:2012mk}).   We readily acknowledge the 
value of the differential distribution in $m_{t \bar{t}}$ for constraints on models, but we defer this 
study to future work.

In addition to the top quark asymmetry, the {\em charged lepton} forward-backward asymmetry 
$\afbl$ is also measured at the Tevatron. The D0 collaboration reports 
$A^{\ell}_{\rm FB}= (11.8 \pm 3.2)\%$, about  2.2 standard deviations 
above the SM value.   In Sec.IV and in the Appendix, we explain the kinematic and 
dynamic aspects of the relationship between the asymmetries $\afbt$ and $\afbl$ based on the 
$(V - A)$ spin correlation between charged leptons and different polarization states of the top quark. 
We show that $\afbl$ and $\afbt$ are strongly positively correlated for {\em right-handed} top quarks.  
For {\em left-handed} top quarks, the strength of the correlation depends on how much the top quark 
is boosted.  Since most of the $t\bar{t}$ events are  produced in the threshold region, the positive 
values of $\afbt$ and $\afbl$ measured at D0 indicate that more right-handed than 
left-handed top quarks are being produced, in contrast to the SM expectation of equal rates.  This is 
a second manifestations of disagreement of asymmetry data with the SM, independent of the discrepancy 
of the magnitude of $\afbt$.   We hasten to remark, however, that the current uncertainties are large.  
The reported D0 data are based on only about half the recorded data set.   Analysis of the full D0 data 
set is desirable, and it would be helpful to have an independent measurement of  $\afbl$ from the CDF 
collaboration.  There is great value in making measurements of both $A_{FB}^t$ and $A_{FB}^\ell$ 
because their correlation can be related through top quark polarization to the underlying 
dynamics of top quark production.   

In Sec. V, we present predictions for the correlation of $A_{FB}^\ell$ with $A_{FB}^t$ at the Tevatron, 
and for the charged lepton asymmetry $A_C^\ell$ with the top quark asymmetry $A_C^t$ at the LHC.   
These predictions are based on the allowed parameter space of the two benchmark new physics models, 
the $W'$ and $G'$ models, determined from our fit to the CDF data on $A_{FB}^t$.   In the case of both 
models, the allowed parameters produce a range of values for the ratios $A_C^\ell/ A_C^t$ at the LHC 
and $A_{FB}^\ell/A_{FB}^t$ at the Tevatron, aligned along approximately straight lines in plots of 
$A_C^\ell$ vs $A_C^t$ and $A_{FB}^\ell$ vs $A_{FB}^t$. Ideally, precise data would provide a 
definite point in the two dimensional plot and tightly constrain the parameter space. The two benchmark 
models we consider are illustrative of the spectrum of possibilities in that the axigluon model produces 
an equal number of right-handed and left-handed top quarks, wheres the flavor-changing $W^\prime$ 
model produces dominantly right-handed top quarks.   

As a final point, we remark that the definitions of the asymmetries require a specification of the 
reference frame in which they are measured, whether the laboratory frame or the $t\bar t$ rest 
frame.  In this paper,  we begin with $\afbt$ in the $t\bar t$ rest frame since the highest 
statistics value of $\afbt$ is measured by CDF in the $t \bar t$ rest frame at the Tevatron. On the 
other hand, the only Tevatron data on $\afbl$ are measured by D0 in the lab frame.  
To take frame dependence into account, one could begin from 
\be
\frac{\afbl(lab)}{\afbt(t\bar t)} = \frac{\afbl(lab)}{\afbt(lab)}\frac{\afbt(lab)}{\afbt(t\bar t)}.
\ee   
The boost tends to reduce $\afbt$ in laboratory frame relative to the $t\bar t$ frame
~\cite{Antunano:2007da}. The reduction is about $30\%$  for the SM, but may be different when 
new physics is included since the kinematics of $t\bar t$ change slightly.  As a result, 
$\afbl(lab)/\afbt(t\bar t)$ will be smaller than  $\afbl(lab)/\afbt(lab)$.   Rather than apply uncertain 
correction factors, we use the D0 laboratory frame data on $\afbl$, but we urge the D0 
collaboration to measure their $\afbt$ in the laboratory frame in order to have a more transparent 
comparison with new physics predictions. A better comparison with theoretical expectations of the 
correlation between the charged lepton asymmetry and top quark asymmetry would be possible 
with a D0 update of $\afbl$ and $\afbt$ in the same frame with their full data set.  

\begin{acknowledgments}
We thank Dr. Jiang-Hao Yu for his
collaboration and contributions during the early
stages of this research.
The work of E.L.B., C.R.C. and H.Z. at Argonne is supported in part by the U.S.
DOE under Grant No.~DE-AC02-06CH11357. H.Z. is also supported by DOE under Grant 
No. DE-FG02-94ER40840.   
Q.H.C. is supported by the National Natural Science Foundation of China under Grant 
No. 11245003.
\end{acknowledgments}

\newpage

\appendix

\section{Energy and angular distributions of the charged lepton}

We present our detailed calculation of the energy and angular distributions of the charged lepton 
from the decay $t \rightarrow W^+ b \rightarrow b \ell^+ \nu$.  

\subsection{The charged-lepton distributions}

In the top quark rest frame, the energy and angular distribution of the charged lepton $\ell^+$ is
\begin{eqnarray}
\frac{d\Gamma}{dxdz} & = & \frac{\alpha_{W}^{2}m_{t}}{32\pi AB}x(1-x)\frac{1+\lambda_t z}{2}
{\rm Arctan}\left[\frac{Ax}{B-x}\right],
\end{eqnarray}
where $x\equiv2E_{\text{\ensuremath{\ell}}}/m_{t}$ ($E_{\ell}$ is
the energy of the charged lepton) and $z\equiv\cos\theta$.  The angle $\theta$ is  
the angle between the direction of motion of the lepton and the top quark
spin direction, $\lambda_t$ denotes the helicity of the top-quark
($\lambda_t=+$ for a right-handed top-quark while $\lambda_t=-$ for a left-handed top-quark), 
$A=\Gamma_{W}/m_{W}$ and $B=m_{W}^{2}/m_{t}^{2}$.
The function $\rm Arctan$ is defined as 
\begin{eqnarray*}
{\rm Arctan}(x) & = & \begin{cases}
\arctan x, & x\geqslant0,\\
\pi+\arctan x, & x<0.
\end{cases}
\end{eqnarray*}
Taking the narrow width approximation for the $W$, we have
\begin{eqnarray}
\frac{d\Gamma}{dxdz} & = & \frac{\alpha_{W}^{2}m_{t}}{32AB}~x(1-x)\frac{1+\lambda_t z}{2}~\Theta(x-B),
\end{eqnarray}
where $\Theta(x)$ is the Heaviside step function, and 
$\Theta(x-B)$ ensures the top quark decays into an on-shell $W$-boson. 

Note that the energy distribution and the angular distribution are separable functions in the 
top quark rest frame.   This implies that, after an integration over the angular distribution, the 
energy distributions of the leptons are identical from left-handed and right-handed top quarks.  

\subsection{$R_F$ along the direction of motion of a boosted top-quark}

We consider next a boost of the top quark along its helicity axis with a velocity $\beta$.  
As a result of the boost, the angular $z$ and energy $x =2E_{\ell}/E_t$ distributions of the lepton become 
correlated.   

\begin{figure}[!htb]
\includegraphics[scale=0.6]{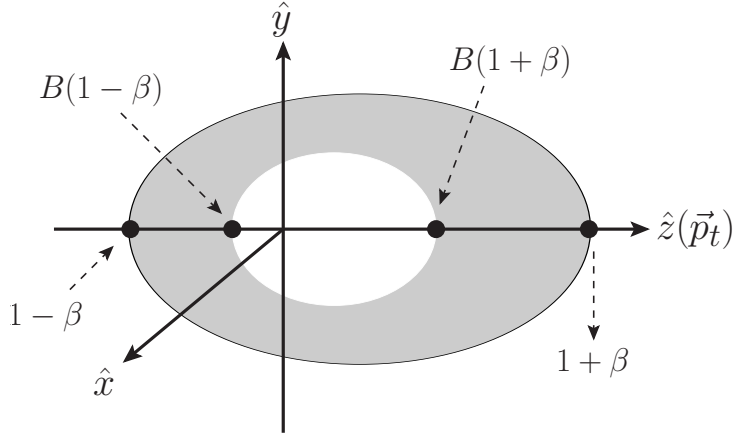}
\caption{Illustration of the charged lepton distribution from a boosted top quark in the laboratory frame. 
The top quark is moving along the $\hat{z}$-axis. The white inner region is not allowed for an 
on-shell $W$-boson. The four black-bold points along the $\hat{z}$ axis show the limits of integration 
of the charged-lepton energy ratio $x$.  }
\label{fig:leptoncone}
\end{figure}

The lepton momentum and angular distribution in this new frame of reference is
\bea
\frac{d\Gamma}{dxdz} & = & \frac{\alpha_{W}^{2}m_{t}}{64AB}\frac{x}{1-\beta^{2}}\left[1-\frac{x}
{1-\beta^{2}}(1-z\beta)\right]\left(1+\lambda_{t}\frac{z-\beta}{1-z\beta}\right)\Theta\left(x-\frac{B(1-\beta^{2})}{1-z\beta}\right).
\label{eq:A3}
\eea
Since the lepton's energy spectrum cannot be negative, the upper limit of the integration over 
$x$ is determined by the following condition
\be
1 - \frac{x}{1-\beta^{2}}(1-z\beta) \ge 0,
\qquad {\rm i.e.}\qquad x_{\rm max}= \frac{1-\beta^2}{1-z\beta}.
\ee
The lower limit is fixed by the Heaviside function in Eq.~\ref{eq:A3},
\be
x_{\rm min}=B\frac{1-\beta^2}{1-z\beta}.
\ee 
Figure~\ref{fig:leptoncone} shows the lepton distribution along the direction of motion of the 
top-quark $\hat{z}(\vec{p}_t)$ in the boosted frame.  The intercepts along the $\hat{z}$-axis 
(i.e. the four black-bold points) are determine by the upper and lower limits of $x$ stated above.  
Only the shaded region is allowed by kinematics, and the inner white region is excluded by the 
on-shell condition of the $W$-boson.  
The angular distribution of the charged lepton is 
\be
\frac{d\Gamma}{dz}= \int^{x_{\rm max}}_{x_{\rm min}}
\frac{d\Gamma}{dz dx} dx = \frac{\alpha_W^2m_t\left(1-B\right)^2\left(1+2B\right)\left(1-\beta^2\right)
\left[1-z\beta+\lambda_t\left(z-\beta\right)\right]}{384AB\left(1-z\beta\right)^3},
\ee
from which we obtain the normalized angular distribution: 
\be
\frac{1}{\Gamma}\frac{d\Gamma}{dz}=\frac{(1-\beta^{2})\left[1-z\beta+\lambda_{t}(z-\beta)\right]}{2(1-z\beta)^{3}}.
\ee

Along the direction of motion of the top quark, the charged lepton is in the forward region with 
$z=\cos\theta > 0$ and in the backward region with $z=\cos\theta<0$. 
The partial width of the charged lepton in the forward region  is 
\bea
\Gamma_{F}&=&\int_{0}^{1}\frac{d\Gamma}{dz}dz
= \frac{\alpha_{W}^{2}m_{t}}{384AB}(1-3B^{2}+2B^{3})(1+\beta)\left[1+\frac{\lambda_{t}}{2}(1-\beta)\right],
\eea
and the partial width of the charged lepton in the backward region is
\bea
\Gamma_{B}&=&\int_{-1}^{0}\frac{d\Gamma}{dz}dz 
=\frac{\alpha_{W}^{2}m_{t}}{384AB}(1-3B^{2}+2B^{3})(1-\beta)\left[1-\frac{\lambda_{t}}{2}(1+\beta)\right].
\eea
The forward fraction ratio $R_F$ is 
\begin{equation}
R_{F}=\frac{\Gamma_F}{\Gamma_F+\Gamma_B}=\frac{1+\beta}{4}\left[2+\lambda_{t}(1-\beta)\right].
\end{equation}
Since $\beta \leq 1$, $R_F$ for a right-handed top quark is always larger than 75\%.   On the other hand, 
for left-handed top quarks, the leptons tend to move opposite the direction of the boost in the top quark 
rest frame.   Owing to this anti-boost effect, there is a critical point of $R_{F}=50\%$ for a left-handed top quark.  
The critical point occurs at $\beta=\sqrt{2}-1\approx0.414$,
i.e. $E_{t}\approx191.5~{\rm GeV}$. 

\subsection{$R_F$ in the laboratory frame}
\label{sub:appRF}

The direction of motion of a top quark does not generally coincide with 
the beam direction, and, therefore, the ratio $R_F$ derived in the previous section 
does not describe the probability of finding a charged lepton 
in the forward region of the detector.   In this section we generalize 
$R_F$ to the situation in which the top quark kinematics in the laboratory frame 
are described by its velocity $\beta$ and rapidity $y_t$, or equivalently, by its 
traverse momentum $p_T$ and rapidity $y_t$.   To obtain $R_F$, we will rotate the 
lepton momentum and angular distribution in Eq.~\ref{eq:A3} to the laboratory frame 
and then integrate over the forward hemisphere in this laboratory frame.   

\begin{figure}[!htb]
\includegraphics[scale=0.5]{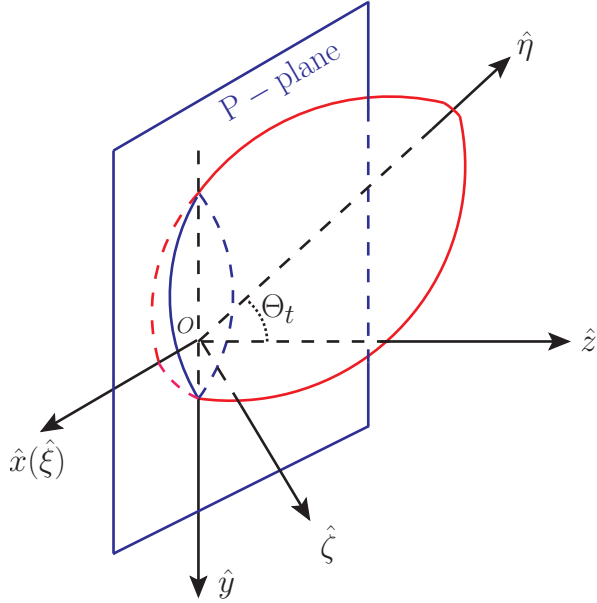}
\caption{Illustration of the charged lepton distribution 
from a boosted top quark in the laboratory frame.
The capital $X,Y,Z$ axes denote the laboratory frame, with the $Z$-axis
being the direction of the beam line.  The top quark is boosted along the $\eta$-axis.}
\label{fig:RF_cone}
\end{figure}

Figure~\ref{fig:RF_cone} illustrates the charged lepton distribution in the laboratory frame 
whose axes are labeled $(\hat{X},\hat{Y},\hat{Z})$.   The top quark boost is along its helicity 
axis $\hat{\eta}$.  The calculation of the decay distribution of the lepton can be carried out 
in the new frame $(\hat{\xi},\hat{\zeta},\hat{\eta})$.  The angle between $\hat{\eta}$ 
and $\hat{Z}$ is denoted $\Theta$, with $k\equiv - \tan\Theta$.  For simplicity we require 
one common transverse direction for the two frames, $\hat{X}$ and $\hat{\xi}$.   The 
important point to make is that the transverse plane $P$ ($\hat{X}$-$\hat{Y}$), 
which separates the forward ($Z>0$) and backward ($Z<0$) regions in the laboratory, 
is not perpendicular to the direction of motion of the top quark.  Our task is to calculate 
the fraction  of the charged leptons that fall in the forward region $Z  > 0$.   

The major semi-axis of the decay ellipsoid is the $\eta$-axis direction, with 
focus at the origin of the two coordinate systems, the top quark decay coordinate 
frame and the laboratory frame.  The $\xi$-axis lies in the 
transverse plane $P$, and the relationship of the values of $\eta$ and $\zeta$ 
for points in this plane is given by the equation of the line obtained by projecting 
the $eta$ and $zeta$ axes onto plane $P$.    
\begin{equation}
\eta=k\zeta,~k<0.
\end{equation}

We introduce polar coordinates,  
\be
\xi=\sin\theta\cos\phi,~\zeta=\sin\theta\sin\phi,~\eta=\cos\theta,
\ee
where $\theta$ is the polar angle, and $\phi$ is the azimuth angle in the frame 
$(\hat{\xi},\hat{\zeta},\hat{\eta})$.  Throughout this work we choose the convention that the $\phi$ 
angle is in the region $[0,2\pi)$, which means $0\leq \phi < 2 \pi$. 
In terms of these polar coordinates, the equation relating points in the plane $P$ becomes 
\begin{equation}
\cot\theta=k\sin\phi.
\end{equation}

Consider the case of a top quark with positive rapidity ($k\leqslant 0$, i.e. $y_t\geqslant 0$). 
Charged leptons on the right (left) of the $P$-plane are in the forward (backward) region in 
the laboratory.  Their momenta satisfy the conditions $\eta>k\zeta$ ($\eta<k\zeta$), respectively. 
In the polar coordinates, the conditions become: 
\bea
&&{\rm forward~region:}~~~\sin\phi>\frac{\cot\theta}{k},\nonumber \\
&&{\rm backward~region:}~\sin\phi<\frac{\cot\theta}{k}.
\eea
These two inequalities then specify the region of integration over the $\theta$ and $\phi$ 
angles as follows: 
\begin{itemize}
\item $\cot\theta/k\leqslant -1$: the condition $\sin\phi > \cot\theta/k$ is always valid. Therefore, 
the charged lepton is always in the forward region.  The integration regions are $\theta\leqslant
{\rm arccot}(-k)$ and $\phi\in(0,2\pi)$.
\item $\cot\theta/k\geqslant 1$ (i.e. $\theta\geqslant{\rm arccot}(k)$): there is no solution 
because no $\phi$ can satisfy $\sin\phi\geq \cot\theta/k \geqslant 1$.  
\item $-1<\cot\theta/k<1$:  
When $\theta>\pi/2$ , 
\be
\phi\in(\arcsin\left(\frac{\cot\theta}{k}\right),\pi-\arcsin\left(\frac{\cot\theta}{k}\right));
\ee
and for $\theta\leqslant\pi/2$, 
\be
\phi\in\biggl[0,\pi-\arcsin\left(\frac{\cot\theta}{k}\right)\biggr)\cup\left(2\pi+\arcsin
\left(\frac{\cot\theta}{k}\right),2\pi\right).
\ee
\end{itemize}
We summarize the integration regions in Table~\ref{tab:int}.  

\begin{table}
\caption{The integration regions of the $\theta$ and $\phi$ angles. Note $k<0$ by definition. }
\label{tab:int}
\begin{tabular}{c|c|c}
\hline 
 & $\cos\theta$ & $\phi$\tabularnewline
\hline  
Part I & $\left(-\frac{k}{\sqrt{1+k^{2}}},1\right)$ & $\left(0,2\pi\right)$\tabularnewline
Part II & $\left(0,-\frac{k}{\sqrt{1+k^{2}}}\right)$ & $\biggl[0,\pi-\arcsin\left(\frac{\cot\theta}
{k}\right)\biggr)\cup\left(2\pi+\arcsin\left(\frac{\cot\theta}{k}\right),2\pi\right)$\tabularnewline
Part III & $\left(\frac{k}{\sqrt{1+k^{2}}},0\right)$ & $\left(\arcsin\left(\frac{\cot\theta}{k}\right),
\pi-\arcsin\left(\frac{\cot\theta}{k}\right)\right)$\tabularnewline
Part IV & $\left(-1,\frac{k}{\sqrt{1+k^{2}}}\right)$ & $\varnothing$\tabularnewline
\hline 
\end{tabular}
\end{table}

The lepton spectrum from decay of the top quark is 
\bea
\frac{d\Gamma}{dxdzd\phi} & = & \frac{\alpha_{W}^{2}m_{t}}{128\pi AB}\frac{x}{1-\beta^{2}}
\left[1-\frac{x(1-z\beta)}{1-\beta^{2}}\right] \nonumber \\
&\times& \left(1+\lambda_{t}\frac{z-\beta}{1-z\beta}\right)\Theta\left(x-\frac{B(1-\beta^{2})}{1-z\beta}\right).
\end{eqnarray}
The partial width for a lepton in the forward region is 
\begin{eqnarray}
\Gamma_{F} & = & \int_{-1}^{1}dz\int_{\frac{B(1-\beta^{2})}{1-z\beta}}^{\frac{(1-\beta^{2})}
{1-z\beta}}dx\int_{\phi_{min}}^{\phi_{max}}\frac{d\Gamma}{dxdzd\phi}d\phi\nonumber \\
 & = & \int_{-\frac{k}{\sqrt{1+k^{2}}}}^{1}dz\int_{\frac{B(1-\beta^{2})}{1-z\beta}}^{\frac{(1-\beta^{2})}
 {1-z\beta}}dx\frac{2\pi d\Gamma}{dxdzd\phi}+\int_{\frac{k}{\sqrt{1+k^{2}}}}^{-\frac{k}
 {\sqrt{1+k^{2}}}}dz\int_{\frac{B(1-\beta^{2})}{1-z\beta}}^{\frac{(1-\beta^{2})}{1-z\beta}}dx
 \frac{d\Gamma}{dxdzd\phi}\left[\pi-2\arcsin\left(\frac{z}{k\sqrt{1-z^{2}}}\right)\right]\nonumber \\
 & = &\underbrace{\int_{-\frac{k}{\sqrt{1+k^{2}}}}^{1}dz\frac{d\Gamma}{dz}}_{\displaystyle\Gamma^{(A)}_F}+
 \underbrace{\int_{\frac{k}{\sqrt{1+k^{2}}}}^{-\frac{k}{\sqrt{1+k^{2}}}}dz\frac{d\Gamma}{2\pi dz}
 \left[\pi-2\arcsin\left(\frac{z}{k\sqrt{1-z^{2}}}\right)\right]}_{\displaystyle \Gamma^{(B)}_F} .
\eea
The values of $\phi_{min}$ and $\phi_{max}$ are listed in the Table, and $\Gamma_{F}^{(A)}$ is 
\be
\frac{\Gamma_{F}^{(A)}}{\Gamma} = \frac{\left(1+\beta\right)}{4\left(\sqrt{1+k^{2}}+k\beta\right)^{2}}
\times\left\{ \lambda_t\left(1-\beta\right)+2\left[1+k^{2}\left(1+\beta\right)+k\sqrt{1+k^{2}}\left(1+\beta\right)\right]\right\} .
\ee
The calculation of $\Gamma_{F}^{(B)}$ involves more steps. 
After integrating $x$, we obtain
\begin{eqnarray}
\frac{\Gamma_{F}^{(B)}}{\Gamma} & = & \int_{\frac{k}{\sqrt{1+k^{2}}}}^{-\frac{k}{\sqrt{1+k^{2}}}}
dz\left[\pi-2\arcsin\left(\frac{z}{k\sqrt{1-z^{2}}}\right)\right]
\frac{\left(1-\beta^{2}\right)}{4\pi\left(1-z\beta\right)^{3}}\left[1-\lambda_t\beta+
\left(\lambda_t-\beta\right)z\right] \nonumber \\
 & = & \int_{-1}^{1}dt\underbrace{\left[-1+\frac{k\beta t}{\sqrt{1+k^{2}t^{2}}}-\lambda_{t}\left(\frac{kt}
 {\sqrt{1+k^{2}t^{2}}}-\beta\right)\right]
\times\frac{k\left(1-\beta^{2}\right)\left(\pi-2\arcsin t\right)}{4\pi\left(\sqrt{1+k^{2}t^{2}}
-k\beta t\right)^{3}}}_{\displaystyle{f(t)}},\nonumber \\
\end{eqnarray}
where we change the integration variable $z$ to $t=z/(k\sqrt{1-z^{2}})$ in the second step. 
The integration can be done analytically, but special care is needed at the upper and lower 
limit where the integral is not analytically continuous. We approach the upper bound from the 
left and the lower limit from the right, obtaining 
\begin{eqnarray}
\lim_{t\to1^{-}}\int f\left(t\right)dt & = & \frac{\beta}{4\sqrt{1+k^{2}\left(1-\beta^{2}\right)}}
+\frac{\lambda_{t}\left(1+k^{2}\right)\left(1-\beta^{2}\right)}{8\left[1+k^{2}\left(1-\beta^{2}\right)
\right]^{3/2}} -\frac{i}{2\pi}\ln\left(k\sqrt{1+k^{2}}\right), \nonumber \\
\lim_{t\to\left(-1\right)^{+}}\int f\left(t\right)dt & = & -\frac{\beta}{4\sqrt{1+k^{2}\left(1-\beta^{2}\right)}}
+\frac{k\sqrt{1+k^{2}}\left(1-\beta^{2}\right)+\beta}{2\left[1+k^{2}\left(1-\beta^{2}\right)\right]}
-\frac{\lambda_{t}\left(1+k^{2}\right)\left(1-\beta^{2}\right)}{8\left[1+k^{2}\left(1-\beta^{2}\right)\right]^{3/2}}\nonumber \\
&+&\frac{\lambda_{t}\left(1-\beta^{2}\right)}{4\left(\sqrt{1+k^{2}}+k\beta\right)^{2}}
 -\frac{i}{2\pi}\ln\left(k\sqrt{1+k^{2}}\right). 
\end{eqnarray}
Hence, $\Gamma_F^{(B)}/\Gamma$ is 
\begin{eqnarray}
\frac{\Gamma_{F}^{(B)}}{\Gamma} & = & \frac{\beta}{2\sqrt{1+k^{2}\left(1-\beta^{2}\right)}}
+\frac{\lambda_{t}\left(1+k^{2}\right)\left(1-\beta^{2}\right)}{4\left[1+k^{2}\left(1-\beta^{2}\right)\right]^{3/2}}
 -\frac{k\sqrt{1+k^{2}}\left(1-\beta^{2}\right)+\beta}{2\left[1+k^{2}\left(1-\beta^{2}\right)\right]} \nonumber \\
&-&\frac{\lambda_{t}\left(1-\beta^{2}\right)}{4\left(\sqrt{1+k^{2}}+k\beta\right)^{2}}.
\end{eqnarray}
Finally, for a top quark in the forward region, i.e. $k\leqslant 0$ or $y_{t}\geqslant 0 $, the 
fraction of leptons in the forward region is 
\begin{eqnarray}
R_{F} \equiv \frac{\Gamma_F}{\Gamma}=  \frac{1}{2}+\frac{\beta}{2\sqrt{1+k^{2}-k^{2}\beta^{2}}}
+\frac{\lambda_{t}\left(1+k^{2}\right)\left(1-\beta^{2}\right)}{4\left(1+k^{2}-k^{2}\beta^{2}\right)^{3/2}}, 
\end{eqnarray}
For a top quark in the backward region the result is $1-R_{F}$ (choosing the opposite $k$). 

We may use $k^{2}=\tan^{2}\Theta_t=-1+\beta^{2}\coth^{2}y_{t}$ to make the connection to the 
top quark rapidity more apparent:
\begin{equation}
R_{F}^{\lambda_{t}}\left(\beta,y_{t}\right)=\begin{cases}
\displaystyle\frac{1}{2}+\frac{1}{2\left(1+\gamma^{-2}\coth^{2}y_{t}\right)^{1/2}}+\frac{\lambda_{t}
\coth^{2}y_{t}}{4\beta\gamma^{2}\left(1+\gamma^{-2}\coth^{2}y_{t}\right)^{3/2}}, & y_{t}\geqslant0,\\
\displaystyle\frac{1}{2}-\frac{1}{2\left(1+\gamma^{-2}\coth^{2}y_{t}\right)^{1/2}}-\frac{\lambda_{t}
\coth^{2}y_{t}}{4\beta\gamma^{2}\left(1+\gamma^{-2}\coth^{2}y_{t}\right)^{3/2}}, & y_{t}<0.
\end{cases}
\end{equation}
One could also choose $p_T$ (the transverse momentum of the top-quark) and $y_t$ as 
the independent kinematic variables. Using the kinematic relations
\be
k = -\frac{p_{T}\csch y_t}{\sqrt{m_{t}^{2}+p_{T}^{2}}},
\qquad 
\beta=\sqrt{1-\frac{m_t^2}{m_t^2+p_T^2}\text{sech}^2 y_t}
\ee
one can rewrite the $R_{F}$ as:
\begin{equation}
R_{F}^{\lambda_{t}}\left(p_{T},y_{t}\right)=\begin{cases}
\displaystyle\frac{1}{2}+\frac{1}{2}\left(1+\frac{m_{t}^{2}\text{csch}^{2}y_{t}}{m_{t}^{2}+
p_{T}^{2}}\right)^{-1/2}+\frac{\lambda_{t}}{4}\left(\frac{m_{t}^{2}\text{csch}^{2}y_{t}}{m_{t}^{2}+p_{T}^{2}}\right)\\
\displaystyle\qquad
\times\left(1+\frac{m_{t}^{2}\text{csch}^{2}y_{t}}{m_{t}^{2}+p_{T}^{2}}\right)^{-3/2}
\left(1-\frac{m_{t}^{2}\text{sech}^{2}y_{t}}{m_{t}^{2}+p_{T}^{2}}\right)^{-1/2}, & y_{t}\geqslant0,\\ 
\nonumber \\
\displaystyle\frac{1}{2}-\frac{1}{2}\left(1+\frac{m_{t}^{2}\text{csch}^{2}y_{t}}{m_{t}^{2}+
p_{T}^{2}}\right)^{-1/2}-\frac{\lambda_{t}}{4}\left(\frac{m_{t}^{2}\text{csch}^{2}y_{t}}{m_{t}^{2}+p_{T}^{2}}\right)\\
\displaystyle\qquad
\times\left(1+\frac{m_{t}^{2}\text{csch}^{2}y_{t}}{m_{t}^{2}+p_{T}^{2}}\right)^{-3/2}
\left(1-\frac{m_{t}^{2}\text{sech}^{2}y_{t}}{m_{t}^{2}+p_{T}^{2}}\right)^{-1/2}, & y_{t}<0.
\end{cases}
\end{equation}

\subsection{Critical Behavior of $R_{F}$}

A few interesting features of the ratio $R_{F}$ are worthy of note.
For left-handed top quarks, when $p_{T}$ is not large,  peak structure is present 
in $R_{F}^{\lambda_{t}}\left(p_{T},y_{t}\right)$,  and there is more than one value of 
$y_{t}$ which satisfies the equation $R_{F}^{\lambda_{t}}\left(\beta,y_{t}\right)=0.5$.

In principle, a peak position can be obtained by solving the equation
\begin{equation}
\frac{\partial R_{F}^{\lambda_{t}}\left(p_{T},y_{t}\right)}{\partial y_{t}}=0, 
\end{equation}
The derivative is not amenable to an analytic solution, but we can still determine the critical 
value of $p_{T}^{c}$.   
When $p_{T}>p_{T}^{c}$, there is no peak structure in $R_{F}^{\lambda_{t}}\left(p_{T},y_{t}\right)$.   
When $p_{T}<p_{T}^{c}$, $R_{F}^{\lambda_{t}}\left(p_{T},y_{t}\right)$ shows peak structure for 
left-handed top quarks.   Solving 
\begin{equation}
\frac{\partial R_{F}^{\lambda_{t}}\left(p_{T}^{c},y_{t}\right)}{\partial y_{t}}\bigg|_{y_{t}=0}=0,
\end{equation}
to obtain $p_{T}^{c}$, we find  
\begin{equation}
\frac{m_{t}^{2}+\left(p_{T}^{c}\right)^{2}-2p_{T}^{c}\sqrt{m_{t}^{2}+\left(p_{T}^{c}\right)^{2}}}{4m_{t}p_{T}^{c}}=0.
\end{equation}
The only physical solution is $p_{T}^{c}=m_{t}/\sqrt{3}\simeq100\text{GeV}$.

For left-handed top quarks, there are values of the boost for which the equation 
$R_{F}^{\lambda_{t}}\left(\beta,y_{t}\right)=0.5$ has more than one solution.   In this
interval of $\beta$, $R_{F}^{\lambda_{t}}\left(\beta,y_{t}\right)$ is not far from $0.5$ 
and is nearly constant.   The solution of $R_{F}^{\lambda_{t}}\left(\beta,y_{t}\right)=0.5$
is $\coth^{2}y_{t}=-\gamma^{2}\left(1+\frac{\lambda_{t}}{2\beta}\right)^{-1}$.
Because $k^{2}=-1+\beta^{2}\coth^{2}y_{t}$ should be greater than 0,  we require 
$\coth^{2}y_{t}\geqslant1/\beta^{2}$.  In the region 
$\beta\in\left(\sqrt{2}-1,{\displaystyle \frac{1}{2}}\right)$ we find that 
$R_{F}^{\lambda_{t}}\left(\beta,y_{t}\right)$ is nearly constant for left-handed top quartks.   
The corresponding values of the energy
of the top quark are $E_{t}\in\left(191.5\text{GeV},{\displaystyle 201.3\text{GeV}}\right)$.

\bibliographystyle{apsrev}
\bibliography{newdraft5.27}
  
\end{document}